\documentclass[a4paper]{article}

\usepackage{amsmath, amssymb, amsthm}
\usepackage{lineno,hyperref}
\usepackage{enumerate}

\usepackage{lineno,hyperref}
\usepackage{comment}

\usepackage{mathdots}
\usepackage{enumerate}

\usepackage[all]{xy}

\usepackage{amsthm}

\newtheorem{thm}{Theorem}[section]

\newtheorem{cor}[thm]{Corollary}

\newtheorem{prop}[thm]{Proposition}
\newtheorem{lem}{Lemma}[section]

\newtheorem{Def}{Definition}[section]
\newtheorem{ex}{Example}

\allowdisplaybreaks

\begin{document}
	
		\title{The tt*-structure for the quantum cohomology of complex Grassmannian}
		\author{Tadashi Udagawa}
	    \date{}
	    \maketitle
		
		\begin{abstract}
		The tt*-equation (topological-anti-topological fusion equation) was introduced by S. Cecotti and C. Vafa for describing massive deformation of supersymmetric conformal field theories. B. Dubrovin formulated the tt*-equation as a flat bundle, called tt*-structure. In this paper, we construct a tt*-structure for the quantum cohomology of the Grassmannian of complex \(k\)-plane and obtain global solutions to the tt*-equation, following the idea of Bourdeau. We give a precise mathematical formulation and a description of the solutions by using p.d.e. theory and the harmonic map theory developed by J. Dorfmeister, F. Pedit and H. Wu (the DPW method). Furthermore, we give an isomorphism between tt*-structure for the \(k\)-th exterior product of tt*-structure for the quantum cohomology of the complex projective space and the tt*-structure for the quantum cohomology of the Grassmannian.
		\end{abstract}
		\vspace{10pt}

    {\flushleft{{\it Keywords:} quantum cohomology, tt*-Toda equation, DPW method}

	
\section{Introduction}
In 1991 \cite{CV1991}, S. Cecotti and C. Vafa introduced the topological anti-topological fusion (tt*)-equations to describe a deformation of \(N=2\) supersymmetric field theories. In mathematics \cite{D1993}, B. Dubrovin formulated the tt*-equation as the flatness condition on a flat bundle (called tt*-structure). and showed that solutions to the tt*-equations correspond to harmonic maps into the symmetric space \({\rm GL}_n \mathbb{R}/{\rm O}_n\). Much later, M. Guest, A. Its and C-S. Lin found all global solutions to the ``Toda-type'' of tt*-equation (tt*-Toda equation) by using p.d.e. theory \cite{GIL20151} and isomonodromy theory \cite{GIL20152}, \cite{GIL2020}. The tt*-Toda equation was motivated by a deformation of the (small) quantum cohomology \(qH^*({\rm Gr}(k,\mathbb{C}^{k+N}))\) of the Grassmannian \({\rm Gr}(k,\mathbb{C}^{k+N})\) \cite{B1995}, \cite{CV1991}. In physics \cite{B1995}, M. Bourdeau studied the Grassmannian \(\sigma\)-model and constructed a solution to the tt*-equation consisting of the tt*-Toda equations.\vskip\baselineskip

The purpose of this paper is to construct a tt*-structure for \(qH^*({\rm Gr}(k,\mathbb{C}^{k+N}))\) and to characterize its solutions using p.d.e. theory and harmonic map theory. We give an explicit description of the tt*-equation for \(qH^*({\rm Gr}(k,\mathbb{C}^{k+N}))\) and we show that solution can be constructed as the ``\(k\)-th exterior product'' of solutions to the tt*-Toda equation. In the language of conformal field theory \cite{B1995}, Bourdeau derived the tt*-equation for \(qH^*({\rm Gr}(k,\mathbb{C}^{k+N}))\). In mathematics, \cite{CDG2019}, G. Cotti, B. Dubrovin and D. Guezzeti investigated an isomonodromic aspects of the tt*-equations for \(qH^*({\rm Gr}(k,\mathbb{C}^{k+N}))\) and proved that its monodromy/Stokes data consists of those for \(qH^*(\mathbb{C}P^{k+N-1})\) case. In 2021 \cite{G2021}, Guest described the tt*-equation for \(qH^*({\rm Gr}(k,\mathbb{C}^{k+N}))\) Lie-theoretically and explain its relation to the Satake correspondence. We give a mathematical formulation of the tt*-equation for \(qH^*({\rm Gr}(k,\mathbb{C}^{k+N}))\) due to the idea of Bourdeau and we characterize the tt*-equation from another point of view (p.d.e theory, harmonic map theory).\vskip\baselineskip

Our first main result is to give a precise mathematical formulation of the tt*-structure constructed from the algebraic structure of \(qH^*({\rm Gr}(k,\mathbb{C}^{k+N}))\) by using the Landau-Ginzburg theory (Proposition 3.3). We also give an explicit description of the tt*-equation and its solution. Our second result is to describe the solution by using solutions to the tt*-Toda equation (Proposition \ref{prop4.2}) and we characterize the solution by the asymptotic behaviour at the origin. Moreover, we give an one-to-one correspondence between solutions to the tt* equation for \(qH^*({\rm Gr}(k,\mathbb{C}^{k+N}))\) and a certain polytope in \(\mathbb{R}^{n+1}\), analogue to the result of Guest, Otofuji \cite{GO2022} (Corollary \ref{thm4.4}).\vskip\baselineskip

Our third result is to give the ``holomorphic data'' associated with the solutions (Proposition 4.5). This data can be interpreted as the chiral data in conformal field theory. In mathematics, holomorphic data is called the generalized Weierstrass data, or DPW data \cite{DPW1998}. This data was not considered by Dubrovin and Cecotti-Vafa, but it is well-known to differential geometers in the context of harmonic map theory and it should play a key role in describing solutions.\vskip\baselineskip

Our fourth result is to show that the tt*-equation for \(qH^*(\mathbb{C}P^{k+N-1})\) induces a tt*-equation for the \(k\)-th exterior product of \(qH^*(\mathbb{C}P^{k+N-1})\), that is, \(\bigwedge^k qH^*(\mathbb{C}P^{k+N-1})\) (Theorem 5.1). We further prove that the tt*-structure for \(\bigwedge^k qH^*(\mathbb{C}P^{k+N-1})\) is isomorphic to the tt*-structure for the quantum cohomology \(qH^*({\rm Gr}(k,\mathbb{C}^{k+N}))\) of the Grassmannian (Theorem 5.2). Furthermore, we interpret the tt*-structure for \(qH^*(\mathbb{C}P^{k+N-1})\) in terms of the Lie-theoretic description introduced by Guest \cite{G2021}. We describe the tt*-structure as a principal \(G\)-bundle and show that tt*-structures for \(qH^*(\mathbb{C}P^{k+N-1})\) and \(qH^*({\rm Gr}(k,\mathbb{C}^{k+N}))\) are induced by the same principal \(G\)-bundle (Proposition 5.4). This result was briefly noted in Example 3.9 of \cite{G2021}.\vskip\baselineskip

This paper is organized as follows. In section 2, we review the definition of tt*-structure and the isomorphism of tt*-structures following Fan, Lan and Yang \cite{FLY2021}. In section 3, we show that the tt*-equation for \(qH^*(\mathbb{C}P^N)\) gives the tt*-Toda equation. For the case \(qH^*({\rm Gr}(k,\mathbb{C}^{k+N}))\), we give a setting of tt*-structure by using the Landau-Ginzburg theory and describe the tt*-equation for \(qH^*({\rm Gr}(k,\mathbb{C}^{k+N}))\) explicitly. In section 4. we construct a global radial solution to the tt*-equation for  \(qH^*({\rm Gr}(k,\mathbb{C}^{k+N}))\) and we find the corresponding asymptotic data and holomorphic data. In section 5,  we give the tt*-structure for \(\bigwedge^kqH^*(\mathbb{C}P^{k+N-1})\) from the tt*-structure on \(qH^*(\mathbb{C}P^{k+N-1})\) (Theorem \ref{thm4.1}). Furthermore, we show that the tt*-structure for \(qH^*(\bigwedge^k\mathbb{C}P^{k+N-1})\) is isomorphic to the tt*-structure for \(qH^*({\rm Gr}(k,\mathbb{C}^{k+N}))\) (Theorem 5.2). As an application, we explain the relation between the result in section 5.1 and the Lie-theoretic description introduced by Guest \cite{G2021}. We give a Lie-theoretic description of our results by considering the principal \(G\)-bundle.

\section{Preliminaries}
\subsection{tt*-structures}
A tt*-structure is a special case of harmonic bundles and it was introduced by Cecotti and Vafa in Physics.  We review tt*-structures on \(\mathbb{C} \backslash (-\infty,0]\) following Dubrovin \cite{D1993} and Fan, Lan, Yang \cite{FLY2021}.
\begin{Def} \label{def2.1}
	A tt*(-geometry) structure \((E,\eta,g,\Phi)\) over a Riemann surface \(\Sigma\) is a holomorphic vector bundle over \(\Sigma\) with a holomorphic structure \(\overline{\partial}_E \), a holomorphic nondegenerate symmetric bilinear form \(\eta \), a Hermitian metric \(g \) and a holomorphic \({\rm End}(E) \)-valued 1-form \(\Phi \) such that
	\begin{enumerate}
		\item [(a)] \(\Phi \) is self-adjoint with respect to \(\eta \), \vspace{1mm}
		
		\item [(b)] a complex conjugate-linear involution \(\kappa\) on \(E\) is given by \(g(a,b) = \eta(\kappa(a),b) \) for \(a,b \in \Gamma(E) \), i.e. \(\kappa^2 = Id_E \) and \(\kappa(\mu a) = \overline{\mu} a \) for \(\mu \in \mathbb{C},\ a \in \Gamma(E) \), \vspace{1mm}
		
		\item [(c)] a flat connection \(\nabla^{\lambda} \) is given by
		\begin{equation}
			\nabla^{\lambda} = D + \lambda^{-1} \Phi + \lambda \Phi^{\dagger_g },\ \ \ \lambda \in S^1, \nonumber
		\end{equation}
		where \(D = \partial_E^g + \overline{\partial}_E \) is the Chern connection and \(\Phi^{\dagger_g} \) is the adjoint operator of \(\Phi \) with respect to \(g\).
	\end{enumerate}
	Given a tt*-structure \((E,\eta,g,\Phi)\), the flatness condition
	\begin{equation}
		F_D = -\left[\Phi,\Phi^{\dagger_g } \right] = -\left(\Phi \wedge \Phi^{\dagger_g} + \Phi^{\dagger_g} \wedge \Phi\right), \nonumber
	\end{equation}
	is called the tt*-equation, where \(F_D = D^2\) is the curvature of \(D\).
\end{Def}\vskip\baselineskip

In this paper, we use the isomorphism of tt*-structures introduced by Fan, Lan and Yang \cite{FLY2021}.
\begin{Def}\label{def2.2}
	Let \((E_j,\eta_j,g_j.\Phi_j),\ j=1,2\) be two tt*-structures over a Riemann surface \(\Sigma\). A bundle map \(T:E_1 \rightarrow E_2\) of two holomorphic bundles is called an isomorphism from \((E_1,\eta_1,g_1,\Phi_1)\) to \((E_2,\eta_2,g_2,\Phi_2)\) if \(T\) satisfies
	\begin{enumerate}
		\item [(1)] \(\eta_1(a,b) = \eta_2(T(a),T(b))\),
		\item [(2)] \(g_1(a,b) = g_2(T(a),T(b))\),
		\item [(3)] \(T((\Phi_1)_X(a)) = (\Phi_2)_X(T(a))\),
	\end{enumerate}
	for all \(a,b \in \Gamma(E), X \in K_{\Sigma}\).
\end{Def}\vskip\baselineskip

The sinh-Gordon equation is an example of the tt*-equation.
\begin{ex}[The sinh-Gordon equation]\label{ex1}
	Given a solution \(w:\mathbb{C} \backslash (-\infty,0] \rightarrow \mathbb{R}\) to the sinh-Gordon equation
	\begin{equation}
		w_{t\overline{t}} = e^{2w} - e^{-2w}. \nonumber
	\end{equation}
	Let \(E = \mathbb{C} \backslash (-\infty,0] \times \mathbb{C}^2\) be a trivial vector bundle and \(e_0,e_1\) be the standard frame of \(E\). We consider the tt*-structure \((E,\eta,g,\Phi)\) for
	\begin{align}
		&\eta(e_i,e_j) = \delta_{i,1-j},\ \ \ g(e_i,e_j) = e^{(-1)^iw}\delta_{i,j}, \nonumber\\
		&\Phi(e_0,e_1) = (e_0,e_1)\left(\begin{array}{cc}
			0 & 1 \\
			1 & 0
		\end{array}\right)dt,\ \ t \in \mathbb{C}^*. \nonumber
	\end{align}
	Then, \((E,\eta,g,\Phi)\) is a tt*-structure over \(\mathbb{C} \backslash (-\infty,0]\) and the tt*-equation is equivalent to the sinh-Gordon equation. \qed
\end{ex}
In general, the tt*-equation is highly nonlinear, and thus, there are the very few explicitly solvable tt*-equations such as the sinh-Gordon equation \cite{MTW1977} and the tt*-Toda equation \cite{GIL20151}, \cite{GIL20152}, \cite{GIL2020}.

\subsection{The DPW method}
The DPW method is a way to construct harmonic map from a Riemann surface into a symmetric space, that was developed by Dorfmeister, Pedit, Wu \cite{DPW1998}. In this paper, we use the DPW method to characterize a tt*-structure by a Lie algebra-valued 1-form (DPW potential). We refer to \cite{DPW1998}.\vskip\baselineskip

Let \(\Sigma \) be a Riemann surface with local coordinate \(z \), \(G \subset {\rm SL}_n \mathbb{C} \) a semisimple Lie subgroup, \(\sigma \) an involution of \(G\) and \(K\) the identity component of the fixed points set \({\rm Fix}(\sigma)\) of \(\sigma\). We split the complexification \(K^{\mathbb{C}} = KB\) by the Iwasawa factorization and the Lie algebra \(\mathfrak{g}\) by
\begin{equation}
	\frak{g} = {\rm Lie}(G) = \frak{k} \oplus \frak{p}, \nonumber
\end{equation}
where
\begin{equation}\frak{k} = \left\{A\ | \ \sigma(A) = A \right\},\ \ \ \ \ \frak{p} = \{A\ | \ \sigma(A)=-A \}. \nonumber
\end{equation}
Here, we denote the derivative of \(\sigma \) by the same notation \(\sigma\). We call
\begin{equation}
	\xi = \frac{1}{\lambda}\xi_{-1}dz + \sum_{j \ge 0} \xi_j dz \lambda^j \in (\Lambda \frak{g}^{\mathbb{C}})_{\sigma} \otimes \Omega^{1,0}_{\Sigma}, \nonumber
\end{equation}
a DPW potential on \(\Sigma\), where \(\xi_j\ (j \ge -1) \) are holomorphic in \(z \) and \(\xi_{2l} \in \frak{k}^{\mathbb{C}}, \xi_{2l-1} \in \frak{p}^{\mathbb{C}}\ (l \in \mathbb{Z}_{\ge 0})\).\vskip\baselineskip

In the DPW method, we construct a harmonic map from \(\xi \) as follows.
\begin{enumerate}[(1)]
	\item First, we assume that \(\Sigma\) is simply-connected. Then we solve 
	\begin{equation}
		d\phi = \phi \xi, \ \ \phi(z_0)= Id, \nonumber
	\end{equation}
	for some base point \(z_0\) and \(\xi \) a holomorphic potential. Since \(\Sigma\) is simply-connected, by the local existence and uniqueness of ordinary differential equations there exists a matrix solution of this equation globally defined on \(\Sigma \). 
	\vspace{2mm} 
	\item Then we can split \(\phi \) via Iwasawa factorization for \((\Lambda G^{\mathbb{C} })_{\sigma}\) (see \cite{DPW1998}, \cite{PS1986} for details) into a product
	\begin{equation}
		\phi = F \phi_+, \nonumber
	\end{equation}
	on some open neighbourhood \(U \subset \Sigma \) of \(z = z_0 \), where 
	\(F \in (\Lambda G)_{\sigma} \), \(\phi_+ \in (\Lambda^+_B G^{\mathbb{C}})_{\sigma} \).
	\vspace{2mm} 
	\item The map
	\begin{equation}
		\pi \circ F|_{\lambda=1}: U \rightarrow G/K, \nonumber
	\end{equation}
	where \(\pi:G \rightarrow G/K \), is a harmonic map (see section 4 of \cite{DPW1998}).
\end{enumerate}\vskip\baselineskip

In general, all solutions to tt*-equations can be constructed from DPW potentials by using the DPW method. In this paper, we call the DPW potential corresponds to a solution to tt*-equation a holomorphic data of the solution, or equivalently, a holomorphic data of the tt*-structure.

\begin{ex}[The tt*-Toda equation \cite{GIL20152}, \cite{GIL2020}]\label{ex2tt}
	The holomorphic data of global radial solutions to the tt*-Toda equation are given by
	\small
	\begin{equation}
		\xi = \frac{1}{\lambda}\left(\begin{array}{cccc}
			& & & z^{l_0}\\
			z^{l_1} & & & \\
			& \ddots & & \\
			& & z^{l_n} &
		\end{array}\right)dz,\ \ \ \ \ \lambda \in S^1,\ z \in \mathbb{C} \backslash (-\infty,0],\ l_0,\cdots,l_n \in \mathbb{R}_{\ge -1}, \nonumber
	\end{equation}
	\normalsize
	where \(\sum_{j=0}^{n}l_j > -(n+1)\) and \(l_j = l_{n+1-j}\) for \(j=1,\cdots,n\). Let \(G = \{g \in {\rm GL}_{n+1} \mathbb{R}\ | \ \Delta g \Delta = g\}, K = G \cap {\rm O}_n\) and \(\sigma(g) = g^{-t}\ (g \in {\rm GL}_{n+1} \mathbb{C})\), where
	\begin{equation}
		\Delta = \left(\begin{array}{ccc}
			& & 1 \\
			& \iddots & \\
			1 & &
		\end{array}\right). \nonumber
	\end{equation}
	We solve \(L^{-1}dL=\xi\) with a certain initial condition \(\phi(0) = \phi_0 \in \Lambda {\rm SL}_{n+1} \mathbb{C}\) and we split \(\phi = F\phi_+\) near \(z=0\) by the Iwasawa factorization for loop group \cite{PS1986}, where \(F \in (\Lambda G)_{\sigma}\) and \(\phi_+ = {\rm diag}(e^{\frac{u_0}{2}},\cdots,e^{\frac{u_n}{2}}) + \mathcal{O}(\lambda) \in (\Lambda^+ G^{\mathbb{C}})_{\sigma}\). We can choose a suitable initial condition such that the Iwasawa factorization is defined on \(\mathbb{C} \backslash (-\infty,0]\) and \(u_j(t,\overline{t}) = u_j(|t|)\) (Corollary 8.1, Corollary 5.1 of \cite{GIL20152}). We put
	\begin{equation}
		w_j = u_j - \frac{1}{n+1}\left\{-2(n+1)\sum_{a=1}^{j}l_a + (2j+1)\sum_{b=1}^{n}l_b + (2j-n)l_0\right\}\log{|z|}, \nonumber
	\end{equation}
	and \( t = \frac{n+1}{n+1+\sum_{a=0}^{n}l_a} z^{\frac{n+1+\sum_{a=0}^{n}l_a}{n+1}}\), then \(\{w_j\}_{j=0}^{n}\) is a global radial solution to the tt*-Toda equation
	\begin{equation}
		(w_j)_{t\overline{t}} = e^{w_j-w_{j-1}} - e^{w_{j+1}-w_j},\ \ \ \ \ (w_{n+1} = w_0, w_{-1} = w_n) \nonumber
	\end{equation}
	with the condition \(w_j + w_{n-j} = 0\) and the asymptotic behaviour
	\begin{equation}
		w_j \sim \frac{1}{n+1+\sum_{a=0}^{n}l_a}\left\{-2(n+1)\sum_{a=1}^{j}l_a + (2j+1)\sum_{b=1}^{n}l_b + (2j-n)l_0\right\}\log{|t|}, \nonumber
	\end{equation}
	as \(t \rightarrow 0\). Thus, the DPW potential \(\xi\) gives a solution to the tt*-Toda equation.
	\qed
\end{ex}

\section{The tt*-structures for the quantum cohomology ring of the Grassmannian}
In physics (\cite{B1995}, 1995), Bourdeau described a solution to the tt*-equation for the quantum cohomology ring \(qH^*({\rm Gr}(k,\mathbb{C}^{k+N}))\) by using solutions to the tt*-equation for the quantum cohomology ring \(qH^*(\mathbb{C}P^{k+N-1})\). In this section, we construct a tt*-structure obtained from the algebraic structure of \(qH^*({\rm Gr}(k,\mathbb{C}^{k+N}))\) and we give a mathematical formulation of the tt*-equation for \(qH^*({\rm Gr}(k,\mathbb{C}^{k+N}))\). In section 3.1, we review the tt*-structure for \(qH^*(\mathbb{C}P^{k+N-1})\) following Cecotti and Vafa \cite{CV1991}. In this case, the corresponding tt*-equation is the tt*-Toda equation. In section 3.2, we construct a tt*-structure for \(qH^*({\rm Gr}(k,\mathbb{C}^{k+N}))\) by using the Landau-Ginzburg theory and we give the tt*-equation for \(qH^*({\rm Gr}(k,\mathbb{C}^{k+N}))\).

\subsection{A tt*-structure for \(qH^*(\mathbb{C}P^{k + N -1})\)}
Set \(n = k + N - 1\). The (small) quantum cohomology ring \(qH^*(\mathbb{C}P^n)\) is given by
\begin{equation}
	qH^*(\mathbb{C}P^n) = \mathbb{C}[z,X] / <h_{n+1}-z>, \nonumber
\end{equation}
where \(h_{n+1} = X^{n+1}\) (see \cite{DGR2010} for details). Define a function
\begin{equation}
	W_{1,N}(z,X) = \frac{1}{n+2}X^{n+2} - zX, \nonumber
\end{equation}
then we have
\begin{equation}
	qH^*(\mathbb{C}P^n) = \mathbb{C}[z,X]/\left(dW_{1,N}/dX\right). \nonumber
\end{equation}
We denote the equivalence class of a polynomial \(p(z,X)\) by \([p(z,X)]\). We consider a holomorphic vector bundle
\begin{equation}
	E^{\mathbb{C}P}_n = \bigsqcup_{z \in \mathbb{C} \backslash (-\infty,0]}\mathbb{C}[z,X]/\left(dW_{1,N}/dX\right) \rightarrow \mathbb{C} \backslash (-\infty,0] : (z,[p(z,X)]) \mapsto z. \nonumber
\end{equation}
In the Landau-Ginzburg theory, a tt*-structure for \(qH^*(\mathbb{C}P^n)\) can be constructed from \(W_{1,N}\) as follows.
\begin{Def}
	We define a nondegenerate holomorphic bilinear form \(\eta^{\mathbb{C}P}\) on \(E^{\mathbb{C}P}_n\) by the Grothendieck residue
	\footnotesize
	\begin{align}
		\eta^{\mathbb{C}P}((z,[a]),(z,[b])) &= \frac{1}{\left(2\pi\sqrt{-1}\right)^n} \int_{\gamma} \frac{a(X)b(X)}{\frac{dW_{1,N}}{dX}}dX 
		&= \sum_{dW_{1,N}=0}a(X)b(X) \left(\frac{d^2W_{1,N}}{dX^2}\right)^{-1}, \nonumber
	\end{align}
	\normalsize
	and a holomorphic 1-form \(\Phi^{\mathbb{C}P}\) on \(E^{\mathbb{C}P}_n\) by
	\begin{equation}
		\Phi_{z\frac{\partial}{dz}}^{\mathbb{C}P}((z,[a])) = \left(z,\left[(-1)\frac{dW_{1,N}}{dz} \cdot a\right]\right) = (z,\left[X \cdot a\right]). \nonumber
	\end{equation}
\end{Def}\vskip\baselineskip

In this section, we use a special frame \(e=(e_0,\cdots,e_n)\) defined by
\begin{equation}
	e_j : \mathbb{C}^* \rightarrow E^{\mathbb{C}P}_n : z \mapsto\left(z,\left[X^j\right]\right),\ \ \ j=0,\cdots,n. \nonumber
\end{equation}

Here, we assume the ``\(\mathbb{Z}_{n+1}\)-symmetry'' with respect to \(\{e_j\}_{j=0}^{n}\)
\begin{equation}
	g^{\mathbb{C}P}(e_i,e_j) = e^{u_j}\delta_{ij}\ \ \ \ \ {\it for\ some}\ u_j:\mathbb{C} \rightarrow \mathbb{R}. \nonumber
\end{equation}
Then, \((E^{\mathbb{C}P}_n,\eta^{\mathbb{C}P},g^{\mathbb{C}P},\Phi^{\mathbb{C}P})\) is a tt*-structure whose tt*-equation is the tt*-Toda equation.

\begin{prop}
	\((E^{\mathbb{C}P}_n,\eta^{\mathbb{C}P},g^{\mathbb{C}P},\Phi^{\mathbb{C}P})\) is a tt*-structure over \(\mathbb{C} \backslash (-\infty,0]\) if and only if \(\{u_j\}_{j=0}^{n}\) satisfies
	\begin{equation}
		\left\{\begin{array}{ll}
			(u_0)_{z\overline{z}} = e^{u_0-u_n} - |z|^{-2}e-{u_1-u_0}, & \\
			(u_j)_{z\overline{z}} = |z|^{-2}e^{u_j-u_{j-1}} - |z|^{-2}e^{u_{j+1}-u_j}, & j=1,\cdots,n-1,\\
			(u_n)_{t\overline{z}} = |z|^{-2}e^{u_n-u_{n-1}} - e^{u_0-u_n},
		\end{array}
		\right. \nonumber
	\end{equation}
	with the condition \(u_j + u_{n-j}=0\) for all \(j\).
\end{prop}
\begin{proof}
	From the definition, we have
	\begin{equation}
		\eta^{\mathbb{C}P}(e_i,e_j) = \sum_{X^{n+1}=z} \frac{X^{i+j-n}}{n+1} = \delta_{i,n-j}. \nonumber
	\end{equation}
	Let \(\kappa^{\mathbb{C}P}\) be a complex conjugate-linear map defined by \(g^{\mathbb{C}P}(a,b) = \eta^{\mathbb{C}P}(\kappa^{\mathbb{C}P}(a),b)\), then \(\kappa^{\mathbb{C}P}(e_j) = e^{u_j}e_{n-j}\). Thus, the condition \(\kappa^2=Id_E\) is equivalent to \(u_j+u_{n-j}=0\) for \(j=0,\cdots,n\).
	Let
	\begin{equation}
		\nabla^{\mathbb{C}P} = \partial_{E^{\mathbb{C}P}_n}^{\dagger_{g^{\mathbb{C}P}}} + \overline{\partial}_{E^{\mathbb{C}P}_n} + \lambda^{-1} \Phi^{\mathbb{C}P} + \lambda \left(\Phi^{\mathbb{C}P}\right)^{\dagger_{g^{\mathbb{C}P}}}, \nonumber
	\end{equation}
	where \(\overline{\partial}_{E^{\mathbb{C}P}_n}\) is the holomorphic structure on \(E^{\mathbb{C}P}_n\), and \(\nabla^{\mathbb{C}P}e = e \cdot \alpha\). Since \(dW_{1,N}/dX=X^{n+1}-z\) and \(X^{n+1}=z\), we have
	\scriptsize
	\begin{align}
		\text{\normalsize \(\alpha\)} &= \left(\begin{array}{ccc}
			(u_0)_z & & \\
			& \ddots & \\
			& & (u_n)_z
		\end{array}\right)dz + \frac{1}{\lambda}\left(\begin{array}{cccc}
			& & & z \\
			1 & & & \\
			& \ddots & & \\
			& & 1 & 
		\end{array}\right)\frac{dz}{z} \nonumber\\
		&\hspace{3cm} 
		+ \lambda \left(\begin{array}{cccc}
			& e^{u_1-u_0} & & \\
			& & \ddots & \\
			& & & e^{u_n-u_{n-1}}\\
			\overline{z}e^{u_0-u_n}& & &
		\end{array}\right)\frac{d\overline{z}}{\overline{z}}. \nonumber
	\end{align}
	\normalsize
	Thus, \((\nabla^{\mathbb{C}P})^2=0\) is equivalent to the system of differential equations stated above.
\end{proof}\vskip\baselineskip

Here, we put
\begin{equation}
	w_j = u_j - \frac{2j-n}{n+1}\log{|z|},\ \ \ j=0,\cdots,n, \nonumber
\end{equation}
and \(t = (n+1)z^{\frac{1}{n+1}}\) then, the tt*-equation for \((E^{\mathbb{C}P}_n,\eta^{\mathbb{C}P},g^{\mathbb{C}P},\Phi^{\mathbb{C}P})\) gives the tt*-Toda equation
\begin{equation}
	(w_j)_{t\overline{t}} = e^{w_j-w_{j-1}} - e^{w_{j+1}-w_j},\ \ \ j=0,\cdots,n, \nonumber
\end{equation}
with the anti-symmetry condition \(w_j + w_{n-j}=0\). Hence, we obtain a tt*-structure on \(E^{\mathbb{C}P}_n\) whose tt*-equation is the tt*-Toda equation.

\subsection{A tt*-structure for \(qH^*({\rm Gr}(k,\mathbb{C}^{k+N}))\)}
We generalize the construction in section 3.1 to the case \(qH^+({\rm Gr}(k,\mathbb{C}^{k+N}))\). Set \(n = k+N-1\). For \(N \ge 2\), the small quantum cohomology of the Grassmannian is given by
\begin{equation}
	qH^*({\rm Gr}(k,\mathbb{C}^{n+1})) = \mathbb{C}[z,X_1,\cdots,X_k] / <h_{N+1},\cdots,h_n,h_{n+1}+(-1)^kz>, \nonumber
\end{equation}
where \(X_r\ (1 \le r \le k)\) are the \(r\)-th elementary symmetric polynomials and \(h_j\ (j \ge 1)\) are the \(j\)-th complete symmetric polynomials in \(k\) variables (see \cite{B2003}, \cite{CDG2019} for details). Define a function
\begin{equation}
	W_{k,N}(z,X_1,\cdots,X_k) = \frac{1}{n+2}p_{n+2}(X_1,\cdots,X_k,0,\cdots,0) + (-1)^k zX_1, \nonumber
\end{equation}
where
\begin{equation}
	p_{n+2}(X_1(t_1,\cdots,t_k),\cdots,X_{n+2}(t_1,\cdots,t_k)) = t_1^{n+2} + \cdots + t_k^{n+2}. \nonumber
\end{equation}
First, we show that \(qH^*({\rm Gr}(k,\mathbb{C}^{n+1}))\) is isomorphic to the fusion ring generated by the function \(W_{k,N}\). We use the result of Gepner \cite{G1991} about the elementary symmetric polynomials.
\begin{lem}[Gepner \cite{G1991}]\label{lem3.2}
	We have
	\begin{equation}
		\frac{(-1)^{j-1}}{j+r} \frac{\partial p_{j+r}}{\partial X_j} = 
		h_r, \nonumber
	\end{equation}
	where \(1 \le j \le k, 0 \le r, 1 \le j+r \le n+2\).
\end{lem}\vskip\baselineskip

From Lemma \ref{lem3.2}, we see that the quantum cohomology \(qH^*({\rm Gr}(k,\mathbb{C}^{n+1}))\) of the Grassmannian can be described by using \(W_{N,k}\).
\begin{prop}
	We have
	\begin{equation}
		qH^*({\rm Gr}(k,\mathbb{C}^{n+1})) = \mathbb{C}[z,X_1,\cdots,X_k] / \left(\frac{\partial W_{k,N}}{\partial X_k},\cdots,\frac{\partial W_{k,N}}{\partial X_1}\right). \nonumber
	\end{equation}
\end{prop}
\begin{proof}
	It follows from the description of \(qH^*({\rm Gr}(k,\mathbb{C}^{n+1}))\) above and Lemma \ref{lem3.2}.
\end{proof}\vskip\baselineskip

We consider a holomorphic vector bundle
\small
\begin{align}
	&E^{\rm Gr}_{k,N} = \bigsqcup_{z \in \mathbb{C} \backslash (-\infty,0]}\mathbb{C}[(-1)^kz,X_1,\cdots,X_k] / \left(\frac{\partial W_{k,N}}{\partial X_k},\cdots,\frac{\partial W_{k,N}}{\partial X_1}\right) \rightarrow \mathbb{C} \backslash (-\infty,0] \nonumber\\
	&\hspace{6cm} 
	: (z,[p(z,X_1,\cdots,X_k)]) \mapsto z. \nonumber
\end{align}
\normalsize
Then, we construct a tt*-structure from the function \(W_{k,N}\) in the same way as the \(qH^*(\mathbb{C}P^n)\) case.
\begin{Def}
	We define a nondegenerate holomorphic bilinear form \(\eta^{\rm Gr}\) on \(E^{\rm Gr}_{k,N}\) by the Grothendieck residue
	\small
	\begin{align}
		\eta^{\rm Gr}((z,[a]),(z,[b])) 
		&= \frac{1}{\left(2\pi\sqrt{-1}\right)^{\binom{k+N}{k} }} \int_{\gamma} \frac{a(X_1,\cdots,X_k)b(X_1,\cdots,X_k)}{\frac{dW_{k,N}}{dX_1} \cdots \frac{dW_{k,N}}{dX_k}}dX_1 \cdots dX_k \nonumber\\
		&= \sum_{dW_{k,N}=0}a(X_1,\cdots,X_k)b(X_1,\cdots,X_k) \left({\rm det}\left(\frac{\partial^2 W_{k,N}}{\partial X_i \partial X_j}\right)\right)^{-1}, \nonumber
	\end{align}
	\normalsize
	and a holomorphic 1-form \(\Phi^{\rm Gr}\) on \(E^{\rm Gr}_{k,N}\) by
	\begin{equation}
		\Phi^{\rm Gr}_{z\frac{d}{dz}}((z,[a])) = \left(z,\left[(-1)^k\frac{dW_{k,N}}{dz} \cdot a\right]\right) = (z,\left[X_1 \cdot a\right]). \nonumber
	\end{equation}
\end{Def}\vskip\baselineskip

We define a holomorphic frame of \(E^{\rm Gr}_{k,N}\) by
\begin{equation}
	e_{\mu_1,\cdots,\mu_k}: \mathbb{C} \backslash (-\infty,0] \rightarrow E^{\rm Gr}_{k,N} : z \mapsto \left(z,\left[s_{\mu_1-k+1, \mu_2-k+2, \cdots, \mu_k}(X_1,\cdots,X_k)\right]\right), \nonumber
\end{equation}
for \(n \ge \mu_1 > \cdots > \mu_k \ge 0\), where \(s_{\mu_1-k+1, \mu_2-k+2, \cdots, \mu_k}\) is the Schur polynomial
\begin{align}
	&s_{\mu_1-k+1, \mu_2-k+2, \cdots, \mu_k}(X_1(t_1,\cdots,t_k),
	\cdots,X_k(t_1,\cdots,t_k)) \nonumber\\
	&\ \ \ \ \ \ \ \ \ \ \ = \frac{1}{\prod_{1 \le i < j \le k}(t_i - t_j)}{\rm det}\left(\begin{array}{cccc}
		t_1^{\mu_1} & t_1^{\mu_2} & \cdots & t_1^{\mu_k} \\
		t_2^{\mu_1} & t_2^{\mu_2} & \cdots & t_2^{\mu_k} \\
		\vdots & \vdots & \ddots & \vdots \\
		t_k^{\mu_1} & t_k^{\mu_2} & \cdots & t_k^{\mu_k}
	\end{array}\right), \nonumber
\end{align}
and \(X_r\ (1 \le r \le k)\) are the \(r\)-th elementary symmetric polynomials
\begin{equation}
	X_r(t_1,\cdots,t_k) = \sum_{k \ge \mu_1 > \cdots > \mu_r \ge 1}t_{\mu_1} \cdots t_{\mu_r}. \nonumber
\end{equation}
We describe \(s_{\mu_1-k+1, \mu_2-k+2, \cdots, \mu_k}\) as a polynomial in \(X_1,\cdots,X_k\). \vskip\baselineskip

As in the case of \(\mathbb{C}P^n\), we assume the \(\mathbb{Z}_{\binom{n+1}{k}}\)-symmetry with respect to \(\{e_{r_1,\cdots,r_k}\}_{n \ge r_1 > \cdots > r_k \ge 0}\)
\small
\begin{equation}
	g^{\rm Gr}(e_{r_1,\cdots,r_k},e_{l_1,\cdots,l_k}) = e^{u_{r_1,\cdots,r_k}}\delta_{r_1,l_1} \cdots \delta_{r_k,l_k}\ \ \ \ \ {\it for\ some}\ u_{r_1,\cdots,r_k}:\mathbb{C} \rightarrow \mathbb{R}. \nonumber
\end{equation}
\normalsize
We consider the tt*equation for \((E^{\rm Gr}_{k,N},\eta^{\rm Gr},g^{\rm Gr},\Phi^{\rm Gr})\) as follows.

\begin{lem}\label{lem3.4}\mbox{}\\
	\begin{itemize}
		\item [(i)] For \(n \ge r_1 > \cdots > r_k \ge 0,\ n \ge l_1 > \cdots > l_k \ge 0\),
		\begin{equation}
			\eta^{\rm Gr}(e_{r_1,\cdots,r_k},e_{l_1,\cdots,l_k}) = (-1)^{\left[\frac{k}{2}\right]}\delta_{r_1,n-l_k} \cdots \delta_{r_k,n-l_1}. \nonumber
		\end{equation}
		
		\item [(ii)] Let \(\kappa^{\rm Gr}\) be a conjugate-linear map on \(E^{\rm Gr}_{k,N}\) defined by \(g^{\rm Gr}(a,b) = \eta\left(\kappa^{\rm Gr}(a),b\right)\), then 
		\begin{equation}
			\kappa^{\rm Gr}(e_{r_1,\cdots,r_k}) = (-1)^{\left[\frac{k}{2}\right]}e^{u_{r_1,\cdots,r_k}}e_{n-r_k,\cdots,n-r_1}. \nonumber
		\end{equation}
	\end{itemize}
\end{lem}
\begin{proof}
	(i) We regard \(X_r\) as the elementary symmetric polynomial in \(t_1,\cdots,t_k\).
	We have
	\begin{equation}
		\frac{\partial W_{k,N}}{\partial t_i} = t_i^{n+1} + (-1)^kz,\ \ \ \frac{\partial^2 W_{k,N}}{\partial t_i \partial t_j} = (n+1)\delta_{ij} \cdot t_i^n. \nonumber
	\end{equation}
	Since
	\begin{equation}
		{\rm det}\left(\frac{\partial X_r}{\partial t_j}\right) = {\rm det}\left(\sum_{\substack{1 \le i_1 < \cdots < i_{r-1} \le k,\\
				i_a \neq j}} t_{i_1} \cdots t_{i_r}\right) = \prod_{1 \le i \le j \le k}(t_i - t_j), \nonumber
	\end{equation}
	we obtain
	\begin{align}
		\left. {\rm det}\left(\frac{\partial^2 W_{k,N}}{\partial X_i \partial X_j}\right) \right|_{dW_{k,N}=0} &= \left. {\rm det}\left(\frac{\partial t_i}{\partial X_j}\right) {\rm det}\left(\frac{\partial^2 W_{k,N}}{\partial t_i \partial t_j}\right) {\rm det}\left(\frac{\partial t_i}{\partial X_j}\right)\right|_{dW_{k,N}=0} \nonumber\\
		&= (n+1)^k z^k t_1^{-1} \cdots t_k^{-1} \prod_{1 \le i < j \le k}(t_i - t_j)^{-2}. \nonumber
	\end{align}
	From the definition, we have
	\footnotesize
	\begin{align}
		&\eta^{\rm Gr}(e_{r_1,\cdots,r_k},e_{l_1,\cdots,l_k}) = \sum_{dW_{k,N}=0} s_{r_1-k+1,\cdots,r_k} s_{l_1-k+1,\cdots,l_k} \frac{z^{-k} t_1 \cdots t_k}{(n+1)^k}\prod_{1 \le i \le j \le k}(t_i - t_j)^2 \nonumber \\
		&= \frac{(n+1)^{-k}z^{-k}}{k!}\sum_{\substack{t_j^{n+1}= (-1)^{k+1}z \\ t_i \neq t_j\ (i \neq j)}} {\rm det}\left(t_i^{r_j}\right) {\rm det}\left(t_a^{l_b}\right) t_1 \cdots t_k \nonumber\\
		&= \frac{(n+1)^{-k}z^{-k}}{k!} \sum_{\substack{t_j^{n+1} = (-1)^{k+1}z \\ t_i \neq t_j\ (i \neq j)}} \sum_{\sigma, \tau \in \mathfrak{S}_k} {\rm sgn}(\sigma) {\rm sgn}(\tau) t_1^{l_{\tau(1)}+r_{\sigma(1)}+1} \cdots t_k^{l_{\tau(k)}+r_{\sigma(k)}+1} \nonumber\\
		&= (n+1)^{-k}z^{-k} \sum_{\substack{t_j^{n+1} = (-1)^{k+1}z \\ t_i \neq t_j\ (i \neq j)}} \sum_{\tau \in \mathfrak{S}_k} {\rm sgn}(\tau) t_1^{r_1+l_{\tau(1)}+1} \cdots t_k^{r_k+l_{\tau(k)}+1} \nonumber\\
		&= \sum_{\tau \in \mathfrak{S}_k} {\rm sgn}(\tau)\prod_{i=1}^{k}\left(\sum_{t_i^{n+1}= (-1)^{k+1}z} (-1)^{k+1}z^{-1}\frac{t_i^{r_i+l_{\tau(i)}+1}}{n+1}\right)
		\nonumber\\
		&= \sum_{\tau \in \mathfrak{S}_k} {\rm sgn}(\tau)\delta_{r_1,n-l_{\tau(1)}} \cdots \delta_{r_k,n-l_{\tau(k)}}. \nonumber
	\end{align}
	\normalsize
	Since \(n \ge r_1 > \cdots > r_k \ge 0, n \ge l_1 > \cdots > l_k \ge 0\), we obtain (i).\vskip\baselineskip
	
	\noindent (ii) Put \(\kappa^{\rm Gr}(e_{r_1,\cdots,r_k}) = \sum_{n \ge a_1 > \cdots > a_k \ge 0}\kappa^{{\rm Gr}, a_1,\cdots,a_k}_{r_1,\cdots,r_k}e_{a_1,\cdots,a_k}\). From the definition, we have
	\small
	\begin{align}
		e^{u_{r_1,\cdots,r_k}}\delta_{r_1,l_1} \cdots \delta_{r_k,l_k} &= g^{\rm Gr}\left(e_{r_1,\cdots,r_k},e_{l_1,\cdots,l_k}\right) 
		= \eta^{\rm Gr}\left(\kappa^{\rm Gr}(e_{r_1,\cdots,r_k}),e_{l_1,\cdots,l_k}\right) \nonumber\\
		&= \sum_{n \ge l_1 > \cdots > l_k \ge 0}\kappa^{{\rm Gr}, a_1,\cdots,a_k}_{r_1,\cdots,r_k}\eta^{\rm Gr}(e_{a_1,\cdots,a_k},e_{l_1,\cdots,l_k}) \nonumber\\
		&= (-1)^{\left[\frac{k}{2}\right]}\kappa^{{\rm Gr}, n-l_k,\cdots,n-l_1}_{r_1,\cdots,r_k}. \nonumber
	\end{align}
	\normalsize
	Hence, we obtain \(\kappa^{\rm Gr}(e_{r_1,\cdots,r_k}) = (-1)^{\left[\frac{k}{2}\right]}e^{u_{r_1,\cdots,r_k}}e_{n-r_k,\cdots,n-r_1}\).
\end{proof}\vskip\baselineskip

In particular, \(\kappa^{\rm Gr}\) is a real structure on \(E^{\rm Gr}_{k,N}\) if and only if \(\{u_{r_1,\cdots.r_k}\}_{n \ge r_1 > \cdots > r_k \ge 0}\) satisfies
\begin{equation}
	u_{r_1,\cdots,r_k} + u_{n-r_k,\cdots,n-r_1} = 0, \nonumber
\end{equation}
for all \(n \ge r_1 > \cdots > r_k \ge 0\). Next, we describe \(\Phi^{\rm Gr}, \left(\Phi^{\rm Gr}\right)^{\dagger_{g^{\rm Gr}}}\) with respect to \(\{e_{r_1,\cdots,r_k}\}_{n \ge r_1 > \cdots > r_k \ge 0}\).

\begin{lem}\label{lem3.5}
	For \(n \ge r_1 > \cdots > r_k \ge 0\), we have
	\small
	\begin{align}
		&\Phi^{\rm Gr}(e_{r_1,\cdots,r_k}) = z^{-1}\sum_{j=1}^{k}e_{r_1,\cdots,r_j+1,\cdots,r_k}dz, \nonumber\\
		&\left(\Phi^{\rm Gr}\right)^{\dagger_{g^{\rm Gr}}}(e_{r_1,\cdots,r_k}) = \overline{z^{-1}}\sum_{j=1}^{k}e^{u_{r_1,\cdots,r_k} - u_{r_1,\cdots,r_j-1,\cdots,r_k}}e_{r_1,\cdots,r_j-1,\cdots,r_k}d\overline{z}, \nonumber
	\end{align}
	\normalsize
	where
	\begin{align}
		&e_{n+1,r_2,\cdots,r_k} = ze_{r_2,\cdots,r_k,0},\ \ \ \ \ e_{r_1,\cdots,r_{k-1},-1} = z^{-1}e_{n,r_1,\cdots,r_{k-1}}, \nonumber\\
		&e_{r_1,\cdots,r_k} = 0\ \ \ {\rm if}\ \exists i,j\ (i \neq j)\ \ {\rm s.t.}\ \ r_i \le r_j. \nonumber
	\end{align}
\end{lem}
\begin{proof}
	From Murnaghan-Nakayama rule, we have
	\begin{align}
		X_1 \cdot s_{\mu_1-k+1,\mu_2-k+2,\cdots,\mu_k} &= (t_1,\cdots,t_k) \cdot s_{\mu_1-k+1,\mu_2-k+2,\cdots,\mu_k}(t_1,\cdots,t_k) \nonumber\\
		&= \sum_{j=1}^{k}s_{\mu_1-k+1,\mu_2-k+2,\cdots,\mu_j-k+j,\cdots,\mu_k}. \nonumber
	\end{align}
	When \(t_1^{n+1} = \cdots = t_k^{n+1} = (-1)^{k-1}z\), we have \(s_{n+1-k+1,\mu_2-k+2,\cdots,\mu_k} = zs_{\mu_2-k+2,\cdots,\mu_k,0}\). Hence, we obtain \(\Phi^{\rm Gr}(e_{r_1,\cdots,r_k}) = z^{-1}\sum_{j=1}^{k}e_{r_1,\cdots,r_j+1,\cdots,r_k}dz\).\vskip\baselineskip
	
	Put \(\left(\Phi^{\rm Gr}\right)^{\dagger_{g^{\rm Gr}}}(e_{r_1,\cdots,r_k}) = \sum_{n \ge l_1 > \cdots > l_k \ge 0}\Phi^{\dagger, l_1,\cdots,l_k}_{r_1,\cdots,r_k}e_{l_1,\cdots,l_k}\). From the definition, we have
	\begin{align}
		&\overline{\Phi^{\dagger, l_1,\cdots,l_k}_{r_1,\cdots,r_k}}e^{u_{l_1,\cdots,l_k}} 
		= g^{\rm Gr}\left(\left(\Phi^{\rm Gr}\right)^{\dagger_{g^{\rm Gr}}}(e_{r_1,\cdots,r_k}),e_{l_1,\cdots,l_k}\right) \nonumber\\
		&= g^{\rm Gr}\left(e_{r_1,\cdots,r_k},\Phi^{\rm Gr}(e_{l_1,\cdots,l_k})\right) \nonumber\\
		&=  z^{-1}\sum_{j=1}^{k}e^{u_{r_1,\cdots,r_k}}\delta_{r_1,l_1} \cdots \delta_{r_j,l_j+1} \cdots \delta_{r_k,l_k}, \nonumber
	\end{align}
	if \(l_1 < n\), and
	\footnotesize
	\begin{align}
		&\overline{\Phi^{\dagger, l_1,\cdots,l_k}_{r_1,\cdots,r_k}}e^{u_{l_1,\cdots,l_k}} \nonumber\\
		&= e^{u_{r_1,\cdots,r_k}}\delta_{r_1,l_2} \cdots \delta_{r_{k-1},l_k} \delta_{r_k,0} + z^{-1}\sum_{j=2}^{k}e^{u_{r_1,\cdots,r_k}}\delta_{r_1,l_1} \cdots \delta_{r_j,l_j+1} \cdots \delta_{r_k,l_k}, \nonumber
	\end{align}
	\normalsize
	if \(l_1 = n\). Hence, we obtain
	\footnotesize
	\begin{equation}
		\left(\Phi^{\rm Gr}\right)^{\dagger_{g^{\rm Gr}}}(e_{r_1,\cdots,r_k}) = \overline{z^{-1}}\sum_{j=1}^{k}e^{u_{r_1,\cdots,r_j,\cdots,r_k} - u_{r_1,\cdots,r_j-1,\cdots,r_k}}d\overline{z} \cdot e_{r_1,\cdots,r_j-1,\cdots,r_k}. \nonumber
	\end{equation}
	\normalsize
\end{proof}\vskip\baselineskip

From Lemme \ref{lem3.4}, \ref{lem3.5}, we obtain a tt*-equation for \((E^{\rm Gr}_{k,N},\eta^{\rm Gr},g^{\rm Gr},\Phi^{\rm})\).
\begin{prop}
	\((E^{\rm Gr}_{k,N},\eta^{\rm Gr},g^{\rm Gr},\Phi^{\rm})\) is a tt*-structure if and only if \(\{u_{r_1,\cdots,r_k}\}_{n \ge r_1 > \cdots > r_k \ge 0}\) is a solution to
	\scriptsize
	\begin{align}
		&\left(u_{r_1,\cdots,r_k}\right)_{z\overline{z}} \nonumber\\
		&= |z|^{-2}\left\{\sum_{\substack{j=1 \\ r_j - 1 \not\equiv r_{j+1} \\ \mod{n+1}}}^{k}e^{u_{r_1,\cdots,r_k} - u_{r_1,\cdots,r_j-1,\cdots,r_k}} - \sum_{\substack{j=1 \\ r_j + 1 \not\equiv r_{j-1} \\ \mod{n+1}}}^{k}e^{u_{r_1,\cdots,r_j+1,\cdots,r_k} - u_{r_1,\cdots,r_k}}\right\}, \nonumber
	\end{align}
	\normalsize
	with the condition
	\begin{itemize}
		\item [(1)] \(u_{r_1,\cdots,r_k} + u_{n-r_k,\cdots,n-r_1} = 0\),
		
		\item [(2)] \(u_{r_1,\cdots,r_k} - u_{r_1,\cdots,r_j-1,\cdots,r_k} = u_{r_1,\cdots,r_i+1,\cdots,r_k} - u_{r_1,\cdots,r_i+1,\cdots,r_j-1,\cdots,r_k}\) for all \(i,j\ (i \neq j,\ 0 \le i, j \le k)\),
	\end{itemize}
	where \(e^{u_{n+1,r_2,\cdots,r_k}} = |z|^2e^{u_{r_2,\cdots,r_k,0}}\) and \(e^{u_{r_1,\cdots,r_k,-1}} = |z|^{-2}e^{u_{n,r_1,\cdots,r_{k-1}}}\).
\end{prop}
\begin{proof}
	From (ii) of Lemma \ref{lem3.4}, we obtain the condition (1). From the definition, the Chern connection \(D^{\rm Gr} = \overline{\partial}_{E^{\rm Gr}_{k,N}} + \partial_{E^{\rm Gr}_{k,N}}^{\dagger_{g^{\rm Gr}}}\)is given by \(D^{\rm Gr}(e_{r_1,\cdots,r_k}) = \left(u_{r_1,\cdots,r_k}\right)_tdt \cdot e_{r_1,\cdots,r_k}\) and we have
	\begin{equation}
		F_{D^{\rm Gr}}(e_{r_1,\cdots,r_k}) = \overline{\partial}_{E^{\rm Gr}}\left(\partial_{E^{\rm Gr}}^{\dagger_{g^{\rm Gr}}}(e_{r_1,\cdots,r_k})\right) = -\left(u_{r_1,\cdots,r_k}\right)_{t\overline{t}}dz \wedge d\overline{z} \cdot e_{r_1,\cdots,r_k}. \nonumber
	\end{equation} 
	From Lemma \ref{lem3.5}, we have
	\footnotesize
	\begin{align}
		&\Phi^{\rm Gr}\left(\left(\Phi^{\rm Gr}\right)^{\dagger_{g^{\rm Gr}}}(e_{r_1,\cdots,r_k})\right) \nonumber\\
		&= \overline{z^{-1}}\sum_{j=1}^{k}e^{u_{r_1,\cdots,r_k} - u_{r_1,\cdots,r_j-1,\cdots,r_k}}\Phi^{\rm Gr}\left(d\overline{z} \cdot e_{r_1,\cdots,r_j-1,\cdots,r_k}\right) \nonumber\\
		&= |z|^{-2}\left\{\sum_{\substack{j=1 \\ r_j - 1 \not\equiv r_{j+1} \\ \mod{n+1}}}^{k}e^{u_{r_1,\cdots,r_k} - u_{r_1,\cdots,r_j-1,\cdots,r_k}}e_{r_1,\cdots,r_k} \right. \nonumber\\
		&\left. \hspace{2,5cm}+ \sum_{\substack{i,j=1 \\ i \neq j}}^{k}e^{u_{r_1,\cdots,r_k} - u_{r_1,\cdots,r_j-1,\cdots,r_k}}e_{r_1,\cdots,r_i+1,\cdots,r_j-1,\cdots,r_k}\right\}dz \wedge d\overline{z}, \nonumber
	\end{align}
	\normalsize
	and
	\scriptsize
	\begin{align}
		&\text{\small \(\left(\Phi^{\rm Gr}\right)^{\dagger_{g^{\rm Gr}}}\left(\Phi^{\rm Gr}(e_{r_1,\cdots,r_k})\right) 
			= z^{-1}\sum_{j=1}^{k}\left(\Phi^{\rm Gr}\right)^{\dagger_{g^{\rm Gr}}}\left(dz \cdot e_{r_1,\cdots,r_j+1,\cdots,r_k}\right)\)} \nonumber\\
		&= |z|^{-2}\left\{\sum_{\substack{j=1 \\ r_j + 1 \equiv r_{j-1} \\ \mod{n+1}}}^{k}e^{u_{r_1,\cdots,r_j+1,\cdots,r_k} - u_{r_1,\cdots,r_k}}e_{r_1,\cdots,r_k} \right. \nonumber\\
		&\left. \hspace{1cm} + \sum_{\substack{i,j=1 \\ i \neq j}}^{k}e^{u_{r_1,\cdots,r_j+1,\cdots,r_k} - u_{r_1,\cdots,r_j+1,\cdots,r_i-1,\cdots,r_k}}e_{r_1,\cdots,r_j+1,\cdots,r_i-1,\cdots,r_k}\right\}d\overline{z} \wedge dz, \nonumber
	\end{align}
	\normalsize
	where \(r_{k+1} = r_1, r_0 = r_k\). Thus, the tt*-equation \(F_{D^{\rm Gr}}(e_{r_1,\cdots,r_k}) = -\left[\Phi^{\rm Gr},\left(\Phi^{\rm Gr}\right)^{\dagger_{g^{\rm Gr}}}\right](e_{r_1,\cdots,r_k})\) is equivalent to
	\begin{align}
		&(u_{r_1,\cdots,r_k})_{t \overline{t}}dz \wedge d\overline{z} \cdot e_{r_1,\cdots,r_k} \nonumber\\
		&= |z|^{-2}\left\{\sum_{\substack{j=1 \\ r_j + 1 \equiv r_{j-1} \\ \mod{n+1}}}^{k}e^{u_{r_1,\cdots,r_k} - u_{r_1,\cdots,r_j+1,\cdots,r_k}} \right. \nonumber\\
		&\left. \hspace{3cm} - \sum_{\substack{j=1 \\ r_j + 1 \equiv r_{j-1} \\ \mod{n+1}}}^{k}e^{u_{r_1,\cdots,r_k} - u_{r_1,\cdots,r_j+1,\cdots,r_k}}\right\}dz \wedge d\overline{z} \cdot e_{r_1,\cdots,r_k} \nonumber\\
		&+ |z|^{-2}\sum_{\substack{i, j = 1 \\ i \neq j}}\left(e^{u_{r_1,\cdots,r_k} - u_{r_1,\cdots,r_j-1,\cdots,r_k}}  \right. \nonumber\\
		&\left. \hspace{1cm} - e^{u_{r_1,\cdots,r_i+1,\cdots,r_k} - u_{r_1,\cdots,r_i+1,\cdots,r_j-1,\cdots,r_k}}\right)dz \wedge d\overline{z} \cdot e_{r_1,\cdots,r_i+1,\cdots,r_j-1,\cdots,r_k}. \nonumber
	\end{align}
	Hence, we obtain the result stated above.
\end{proof}\vskip\baselineskip

\begin{ex}[The quantum cohomology of \({\rm Gr}(2,\mathbb{C}^4)\)]\label{ex2}
	Let
	\begin{equation}
		W_{2,2}(z,X_1,X_2) = \frac{1}{5}\left(X_1^5 - 5X_1^3X_2 + 5X_1X_2^2\right) + zX_1, \nonumber
	\end{equation}
	then
	\begin{equation}
		E^{\rm Gr}_{2,2} = \bigsqcup_{z \in \mathbb{C}^*}\mathbb{C}[X_1,X_2]/<X_1^3-2X_1X_2,X_1^4-3X_1^2X_2+X_2^2+z>, \nonumber
	\end{equation}
	with the frame \(e = (e_{1,0},e_{2,0},e_{2,1},e_{3,0},e_{3,1},e_{3,2})\), where
	\footnotesize
	\begin{align}
		&e_{1,0}(z) = (z,[1]),\ e_{2,0}(z) = (z,[X_1]),\ e_{2,1}(z) = (z,[X_2]),\ e_{3,0}(z) = (z,[X_1^2 - X_2]), \nonumber\\
		&e_{3,1}(z) = (z,[X_1X_2]),\ e_{3,2}(z) = (z,[X_2^2]). \nonumber
	\end{align}
	\normalsize
	Thus, the non-vanishing elements of \(\eta^{\rm Gr}(e_{l_i,l_j},e_{r_a,r_b})\) are
	\begin{align}
		\eta^{\rm Gr}(e_{1,0},e_{3,2}) &= \eta^{\rm Gr}(e_{2,0},e_{3,1}) = \eta^{\rm Gr}(e_{2,1},e_{2,1}) = \eta^{\rm Gr}(e_{3,0},e_{3,0}) \nonumber\\
		& = \eta^{\rm Gr}(e_{3,1},e_{2,0}) = \eta^{\rm Gr}(e_{3,2},e_{1,0}) \nonumber\\ 
		&= -1, \nonumber
	\end{align}
	\(g^{\rm Gr}(e_{r_1,r_2},e_{l_1,l_2}) = e^{w_{r_1,r_2}}\delta_{r_1,l_1}\delta_{r_2,l_2}\)
	\begin{equation}
		\Phi^{\rm Gr}(e) = e \cdot \left(\begin{array}{cccccc}
			0 & 0 & 0 & 0 & z & 0 \\
			1 & 0 & 0 & 0 & 0 & z \\
			0 & 1 & 0 & 0 & 0 & 0 \\
			0 & 1 & 0 & 0 & 0 & 0 \\
			0 & 0 & 1 & 1 & 0 & 0 \\
			0 & 0 & 0 & 0 & 1 & 0
		\end{array}\right)\frac{dz}{z}, \nonumber
	\end{equation}
	and
	\scriptsize
	\begin{align}
		&\text{\small \(\Phi^{\dagger_{g^{\rm Gr}}}(e)\)} \nonumber\\
		\scriptsize
		&= e \cdot \left(\begin{array}{cccccc}
			0 & e^{u_{2,0} - u_{1,0}} & 0 & 0 & 0 & 0 \\
			0 & 0 & e^{u_{2,1} - u_{2,0}} & e^{u_{3,0} - u_{2,0}} & 0 & 0 \\
			0 & 0 & 0 & 0 & e^{u_{3,1} - u_{2,1}} & 0 \\
			0 & 0 & 0 & 0 & e^{u_{3,1} - u_{3,0}} & 0 \\
			\overline{z}e^{u_{1,0} - u_{3,1}} & 0 & 0 & 0 & 0 & e^{u_{3,2} - u_{3,1}}\\
			0 & \overline{z}e^{u_{2,0} - u_{3,2}} & 0 & 0 & 0 & 0
		\end{array}\right)\frac{d\overline{z}}{\overline{z}}, \nonumber
	\end{align}
	\normalsize
	Thus, the tt*-equation is given by
	\begin{align}
		&(u_{1,0})_{z \overline{z}} = e^{u_{1,0} - u_{3,1}} - e^{u_{2,0} - u_{1,0}}, \nonumber\\
		&(u_{2,0})_{z\overline{z}} = |z|^{-2}e^{u_{2,0} - u_{1,0}} + e^{u_{2,0} - u_{3,2}} - |z|^{-2}e^{u_{2,1} - u_{2,0}} - |z|^{-2}e^{u_{3,0} - u_{2,0}}, \nonumber\\
		&(u_{2,1})_{z\overline{z}} = |z|^{-2}e^{u_{2,1} - u_{2,0}} - |z|^{-2}e^{u_{3,1} - u_{2,1}}, \nonumber\\
		&(u_{3,0})_{z\overline{z}} = |z|^{-2}e^{u_{3,0} - u_{2,0}} - |z|^{-2}e^{u_{3,1} - u_{3,0}}, \nonumber\\
		&(u_{3,1})_{z\overline{z}} = |z|^{-2}e^{u_{3,1} - u_{2,1}} + |z|^{-2}e^{u_{3,1} - u_{3,0}} - e^{u_{1,0} - u_{3,1}} - |z|^{-2}e^{u_{3,2} - u_{3,1}}, \nonumber\\
		&(u_{3,2})_{z\overline{z}} = |z|^{-2}e^{u_{3,2} - u_{3,1}} - e^{u_{2,0} - u_{3,2}}, \nonumber
	\end{align}
	with the condition \(u_{r_1,r_2} + u_{3-r_1,3-r_2} = 0\ (3 \ge r_1 > r_2 \ge 0)\) and
	\begin{align}
		&u_{3,2} - u_{3,1} = u_{2,0} - u_{1,0},\ \ \ u_{2,1} - u_{2,0} = u_{3,1} - u_{3,0}, \nonumber\\
		&u_{3,0} - u_{2,0} = u_{3,1} - u_{2,1},\ \ \ u_{1,0} - u_{3,1} = u_{2,0} - u_{3,2}. \nonumber
	\end{align}
	\qed
\end{ex}\vskip\baselineskip

As in the case of \(\mathbb{C}P^n\), we put \(t = (n+1)z^{\frac{1}{n+1}}\) and
\begin{equation}
	w_{r_1,\cdots,r_k} = e^{u_{r_1,\cdots,r_k}} - \left(2\sum_{j=1}^{k}r_j - nk\right)\log{|t|} + \left(2\sum_{j=1}^{k}r_j - nk\right)\log{(n+1)}, \nonumber
\end{equation}
then \(w_{r_1,\cdots,r_k}\) satisfies
\scriptsize
\begin{align}\label{Gr}
	&(w_{r_1,\cdots,r_k})_{t\overline{t}} \nonumber\\
	&= \sum_{\substack{j=1\\ r_j - 1 \not\equiv r_{j-1} \\ \mod n+1}}^{k}e^{w_{r_1,\cdots,r_j,\cdots,r_k} - w_{r_1,\cdots,r_j-1,\cdots,r_k}} 
	- \sum_{\substack{j=1\\ r_j + 1 \not\equiv r_{j+1} \\ \mod n+1 }}^{k}e^{w_{r_1,\cdots,r_j+1,\cdots,r_k} - w_{r_1,\cdots,r_j,\cdots,r_k}}, \tag{Gr} \nonumber
\end{align}
\normalsize
with the condition
\begin{itemize}
	\item [(I)] \(w_{r_1,\cdots,r_k} + w_{n-r_k,\cdots,n-r_1} = 0\),
	
	\item [(II)] \(w_{r_1,\cdots,r_k} - w_{r_1,\cdots,r_j-1,\cdots,r_k} = w_{r_1,\cdots,r_i+1,\cdots,r_k} - w_{r_1,\cdots,r_i+1,\cdots,r_j-1,\cdots,r_k}\) for all \(i,j\ (i \neq j,\ 0 \le i, j \le k)\),
\end{itemize}
where \(r_0 = r_n, r_{k+1} = r_1\), \(w_{-1,r_2,\cdots,r_k} = w_{r_2,\cdots,r_k,n}\) and \(w_{r_1,\cdots,r_{k-1},n+1} = w_{0,r_1,\cdots,r_{k-1}}\).\vskip\baselineskip

Thus, we obtain a tt*-equation constructed from the algebraic structure of \(qH^*({\rm Gr}(k,\mathbb{C}^{k+N}))\) by using the Landau-Ginzburg theory. In the following section, we give a solution to the tt*-equation and we characterize the solution by asymptotic data and holomorphic data.

\section{The tt*-equation for \(qH^*({\rm Gr}(k,\mathbb{C}^{k+N}))\)}
In \cite{GIL20151}, \cite{GIL20152}, \cite{GIL2020}, Guest, Its and Lin characterize global radial solutions to the tt*-Toda equation by the asymptotic behaviour at \(t=0\) (asymptotic data), the DPW potentials correspond to solutions (holomorphic data) and the monodromy data of certain o.d.e. which specified th tt*-equation (Stokes data). Regarding the tt*-equation for \(qH^*({\rm Gr}(k,\mathbb{C}^{k+N}))\), the Stokes data was investigated by Cotti, Dubrovin, Guzzetti and they found that the Stokes data was given by the exterior product of the Stokes data of the tt*-Toda equation \cite{CDG2019}.\vskip\baselineskip

In this section, we construct a global radial solution to the tt*-equation (\ref{Gr}) and we characterize the solution by asymptotic data and holomorphic data. In section 3.1, we show that a solution to the tt*-equation (\ref{Gr}) can be constructed from solutions to the tt*-Toda equation and we characterized the solution by its asymptotic behaviour at \(t = 0\). In section 3.2, we construct the solution from a DPW potential by using the DPW method and the DPW potential is the holomorphic data of the solution to the tt*-equation (\ref{Gr}) with the condition (I), (II).

\subsection{The asymptotic data}
In \cite{GIL20151}, \cite{GIL20152}, \cite{GIL2020}, Guest, Its and Lin found all radial solutions to the tt*-Toda equation
\begin{equation}
	(w_j)_{t\overline{t}} = e^{w_j-w_{j-1}} - e^{w_{j+1}-w_j},\ \ \ j=0,\cdots,n, \nonumber
\end{equation}
with the condition \(w_j + w_{n-j} = 0\). In \cite{GO2022}, Guest, Otofuji gave an one-to-one correspondence between global radial solutions to the tt*-Toda equation and a convex polytope in \(\mathbb{R}^{n+1}\).
\begin{thm}[\cite{GIL20151}, \cite{GL2014}]\label{thm4.3}
	There is an one-to-one correspondence between global radial solutions \(\{w_j\}_{j=0}^{n}\) to the tt*-Toda equation with the asymptotic behaviour
	\begin{equation}
		w_j \sim -m_j \log{|t|}\ \ \ \ \ {\rm as}\ \ \ t \rightarrow 0, \nonumber
	\end{equation}
	and the convex polytope
	\begin{equation}
		\{m = (m_0,\cdots,m_n)\ | \ m_{j-1} - m_j + 2 \ge 0,\ m_j + m_{n-j} = 0\}. \nonumber
	\end{equation}
\end{thm}\vskip\baselineskip

In this paper, we construct a solution to the tt*-equation (\ref{Gr}) consisting of solutions to the tt*-Toda equation. The condition (II) is equivalent to
\begin{itemize}
	\item [(II')] \(\forall i,j, \exists w^{ij}\ {\rm s.t.}\ w_{r_1,\cdots,r_k} = w_{l_1,\cdots,l_k} + w^{i,j}\) for all \((r_1,\cdots,r_k), (l_1,\cdots.l_k)\ (i \in \{r_1,\cdots,r_k\}, j \in \{l_1,\cdots,l_k\}, \{r_1,\cdots,r_k\} \backslash \{i\} = \{l_1,\cdots,l_k\} \backslash \{j\})\). \nonumber
\end{itemize}
The functions \(\{w_{i,j}\}\) satisfy the following lemma.
\begin{lem}\label{lem4.1}
	We have
	\begin{itemize}
		\item [(1)] \(w^{i,j} = - w^{j,i}\),
		
		\item [(2)] \(w^{i,j} + w^{j,l} = w^{i,l}\),
		
		\item [(3)] \(w^{n-i,n-j} = -w^{i,j}\).
	\end{itemize}
\end{lem}
\begin{proof}
	It follows from the definition of \(\{w^{i,j}\}\).
\end{proof}

The functions \(\{w^{i,j}\}\) give solutions to the tt*-Toda equation.
\begin{prop}\label{prop4.1}
	Let \(w_i = \frac{w^{i,n-i}}{2}\ (i= 0,\cdots,n)\), then \(\{w_i\}_{i=0}^{n}\) satisfy the tt*-Toda equation
	\begin{equation}
		(w_i)_{t\overline{t}} = e^{{w_i - w_{i-1}}} - e^{{w_{i+1} - w_i}}, \ \ \ \ \ i= 0, \cdots,n, \nonumber
	\end{equation}
	with the condition \(w_i + w_{n-i} = 0\), where \(w_{n+1} = w_0\) and \(w_{-1} = w_n\).
\end{prop}
\begin{proof}
	From the definition, we have
	\begin{equation}
		(w_{r_1,\cdots,r_k})_{t\overline{t}} = \sum_{\substack{j=1\\ r_j - 1 \not\equiv r_{j-1} \\ \mod n+1}}^{k}e^{w^{r_j,r_j-1}} - \sum_{\substack{j=1\\ r_j - 1 \not\equiv r_{j-1} \\ \mod n+1}}^{k}e^{w^{r_j+1,r_j}}, \nonumber
	\end{equation}
	and then, from (1), (3) of Lemma \ref{lem4.1} we have
	\begin{align}
		(w_{i,n-i})_{t\overline{t}} &= e^{w^{i,i-1}} - e^{w^{i+1,i}} + e^{w^{n-i,n-i-1}} - e^{w^{n-i+1,n-i}} \nonumber\\
		&= 2e^{w^{i,i-1}} - 2e^{w^{i+1,i}}. \nonumber 
	\end{align}
	From (2), (3) of Lemma \ref{lem4.1}, we have
	\begin{align}
		2w^{i,i-1} &= w^{i,i-1} - w^{n-i,n-i+1} = w^{i,j} + w^{j,i-1} - w^{n-i,j} - w^{j,n-i+1} \nonumber\\
		&= w^{i,j} + w^{j,n-j} - w^{i-1,j} - w^{j.n-i+1} \nonumber\\
		&= w^{i,n-i} - w^{i-1,n-i+1}. \nonumber
	\end{align}
	Similarly, we have \(2w^{i+1,i} = w^{i+1,n-i-1} - w^{i,n-i}\). Hence, we obtain
	\begin{align}
		(w_i)_{t\overline{t}} &= \frac{1}{2}(w^{i,n-i})_{t\overline{t}} = e^{\frac{w^{i,n-i}}{2} - \frac{w^{i-1,n-i+1}}{2}} - e^{\frac{w^{i+1,n-i-1}}{2} - \frac{w^{i,n-i}}{2}} \nonumber\\
		&= e^{{w_j - w_{j-1}}} - e^{{w_{j+1} - w_j}}. \nonumber
	\end{align}
\end{proof}\vskip\baselineskip

Thus, we can describe a solution to the tt*-equation (\ref{Gr}) by using a finite number of solutions to the tt*-Toda equation.
\begin{prop}\label{prop4.2}
	Let \(\{w_{r_1,\cdots,r_k}\}_{n \ge r_1 > \cdots > r_k \ge 0}\) be a solution to the tt*-equation (\ref{Gr}) with the condition (I), (II'), then \(\{w_{r_1,\cdots,r_k}\}_{n \ge r_1 > \cdots > r_k \ge 0}\) can be split into
	\begin{equation}
		w_{r_1,\cdots,r_k} = w_{r_1} + \cdots + w_{r_k},\ \ \ \ \ \ \ \ \ n \ge r_1 > \cdots > r_k \ge 0, \nonumber
	\end{equation}
	where \(\{w_j\}_{j=0}^{n}\) is a solution to the tt*-Toda equation with \(w_j + w_{n-j} = 0\).
\end{prop}
\begin{proof}
	From (I), (II'), we have
	\begin{align}
		w_{r_1,\cdots,r_k} &= w^{r_1,n-r_1} + \cdots + w^{r_k,n-r_k} + w_{n-r_k,\cdots,n-r_1} \nonumber\\
		&= w^{r_1,n-r_1} + \cdots + w^{r_k,n-r_k} - w_{r_1,\cdots,r_k}. \nonumber
	\end{align}
	From Proposition \ref{prop4.1}, we obtain the result stated above.
\end{proof}\vskip\baselineskip

Combining Theorem \ref{thm4.3} and Proposition \ref{prop4.2}, we obtain a global radial solution to the tt*-equation (\ref{Gr}). As a corollary, we consider an one-to-one correspondence between solutions \(\{w_{r_1,\cdots,r_k}\}_{n \ge r_1 > \cdots > r_k \ge 0}\) to the tt*-equation (\ref{Gr}) and a subset of \(\mathbb{R}^{\binom{n+1}{k}}\) analogue to the result of Guest, Otofuji.

\begin{cor}\label{thm4.4}
	For \(\{m_{r_1,\cdots,r_k}\}_{n \ge r_1 > \cdots > r_k \ge 0}\), we consider a condition
	\begin{itemize}
		\item [(i)] \(\forall i,j, \exists m^{ij}\ {\rm s.t.}\ m_{r_1,\cdots,r_k} = m_{l_1,\cdots,l_k} + m^{i,j}\) for all \((r_1,\cdots,r_k), (l_1,\cdots.l_k)\ (i \in \{r_1,\cdots,r_k\}, j \in \{l_1,\cdots,l_k\}, \{r_1,\cdots,r_k\} \backslash \{i\} = \{l_1,\cdots,l_k\} \backslash \{j\})\). \nonumber
	\end{itemize}
	There is an one-to-one correspondence between global radial solutions \(\{w_{r_1,\cdots,r_k}\}_{n \ge r_1 > \cdots > r_k \ge 0}\) to the tt*-equation (\ref{Gr}) on \(\mathbb{C}^*\) with the conditions (I), (II) and the asymptotic behaviour
	\begin{equation}
		w_{r_1,\cdots,r_k} \sim -m_{r_1,\cdots,r_k}\log{|t|}\ \ \ {\rm as}\ \ \ t \rightarrow 0, \nonumber
	\end{equation}
	and a set
	\small
	\begin{equation}
		\left\{m^{\rm Gr} = \{m_{r_1,\cdots,r_k}\}_{n \ge r_1 > \cdots > r_k \ge 0} \in \mathbb{R}^{\binom{n+1}{k}}\ \middle| \ \begin{array}{l}
			\text{\(m_{r_1,\cdots,r_k}\) satisfies (i)}, \\
			\ m^{j-1,j} + 2 \ge 0,\ \forall j = 0, \cdots, n, \\
			\ m_{r_1,\cdots,r_k} + m_{n-r_k,\cdots,n-r_k} = 0
		\end{array}
		\right\}. \nonumber
	\end{equation}
	\normalsize
\end{cor}
\begin{proof}
	It follows from Proposition \ref{prop4.1}, \ref{prop4.2} and Theorem \ref{thm4.3}.
\end{proof}\vskip\baselineskip

\begin{ex}[The tt*-equation (\ref{Gr}) for the case \(N = 2, k=2\)]\label{ex4}
	Let \(n = 3\) and \(\{w_{r_1,r_2}\}_{3 \ge r_1 > r_2 \ge 0}\) be a global radial solution to the tt*-equation (\ref{Gr})
	\begin{align}
		&(w_{1,0})_{t\overline{t}} = e^{w_{1,0} - w_{3,1}} - e^{w_{2,0} - w_{1,0}}, \nonumber\\
		&(w_{2,0})_{t\overline{t}} = e^{w_{2,0} - w_{1,0}} + e^{w_{2,0} - w_{3,2}} - e^{w_{2,1} - w_{2,0}} - e^{w_{3,0} - w_{2,0}}, \nonumber\\
		&(w_{2,1})_{t\overline{t}} = e^{w_{2,1} - w_{2,0}} - e^{w_{3,1} - w_{2,1}}, \nonumber\\
		&(w_{3,0})_{t\overline{t}} = e^{w_{3,0} - w_{2,0}} - e^{w_{3,1} - w_{3,0}}, \nonumber\\
		&(w_{3,1})_{t\overline{t}} = e^{w_{3,1} - w_{2,1}} + e^{w_{3,1} - w_{3,0}} - e^{w_{1,0} - w_{3,1}} - e^{w_{3,2} - w_{3,1}}, \nonumber\\
		&(w_{3,2})_{t\overline{t}} = e^{w_{3,2} - w_{3,1}} - e^{w_{2,0} - w_{3,2}}, \nonumber
	\end{align}
	with the condition \(w_{r_1,r_2} + w_{3-r_1,3-r_2} = 0\ (3 \ge r_1 > r_2 \ge 0)\) and
	\begin{align}
		&w_{3,2} - w_{3,1} = w_{2,0} - w_{1,0},\ \ \ w_{2,1} - w_{2,0} = w_{3,1} - w_{3,0}, \nonumber\\
		&w_{3,0} - w_{2,0} = w_{3,1} - w_{2,1},\ \ \ w_{1,0} - w_{3,1} = w_{2,0} - w_{3,2}. \nonumber
	\end{align}
	The asymptotic behaviour of \(\{w_{r_1,r_2}\}\) is given by
	\begin{equation}
		w_{r_1,r_2} \sim -m^{\rm Gr}_{r_1,r_2}\log{|t|}\ \ \ {\rm as}\ \ t \rightarrow 0, \nonumber
	\end{equation}
	where \(\{m_{r_1,r_2}\}\) satisfies
	\begin{align}
		&m_{3,2} - m_{3,1} = m_{2,0} - m_{1,0} \le 2,\ \ \ m_{2,1} - m_{2,0} = m_{3,1} - m_{3,0} \le 2, \nonumber\\
		&m_{3,0} - m_{2,0} = m_{3,1} - m_{2,1} \le 2,\ \ \ m_{1,0} - m_{3,1} = m_{2,0} - m_{3,2} \le 2, \nonumber
	\end{align}
	and \(m_{r_1,r_2} = -m_{3-r_1,3-r_2}\) for \(3 \ge r_1 > r_2 \ge 0\).
	\qed
\end{ex}\vskip\baselineskip

\subsection{The holomorphic data}
For the tt*-Toda equation, the DPW method can be exploited to obtain some data which characterize the solution of the tt*-Toda equation \cite{DGR2010}, \cite{GIL20151}, \cite{GIL20152}, \cite{GIL2020}. Guest, Its and Lin proved that there is a one-to-one correspondence between the solutions and DPW potentials. In this section, we give a holomorphic data for solutions to the tt*-equation (\ref{Gr}).\vskip\baselineskip

The holomorphic data of global radial solutions to the tt*-Toda equation are given by
\begin{equation}
	\xi = \frac{1}{\lambda}\left(\begin{array}{cccc}
		& & & z^{l_0}\\
		z^{l_1} & & & \\
		& \ddots & & \\
		& & z^{l_n} &
	\end{array}\right)dz,\ \ \ \ \ \lambda \in S^1,\ z \in \mathbb{C} \backslash (-\infty,0],\ l_0,\cdots,l_n \in \mathbb{R}_{\ge -1}, \nonumber
\end{equation}
where \(l_j = l_{n+1-j}\) for \(j=1,\cdots,n\). We extend this correspondence to global radial solutions to the tt*-Toda equation (\ref{Gr}). Let \(\phi\) be a solution to \(d\phi = \phi\xi\) in Example \ref{ex2tt}, i.e. \(\phi\) admits an global Iwasawa factorization \(\phi = F\phi_+\) on 
\(\mathbb{C} \backslash (-\infty,0]\).
\vskip\baselineskip

Put \(\xi^{\rm Gr} = \left(\lambda^{-1}\xi^{{\rm Gr}\ b_1,\cdots,b_k}_{a_1,\cdots,a_k}\right)_{\substack{n \ge r_1 > \cdots > r_k \ge 0 \\ n \ge b_1 > \cdots > b_k \ge 0}}dz \in (\Lambda {\mathfrak{sl}}_{\binom{n}{k}} \mathbb{C})_{\sigma} \otimes \Omega^{1,0}_{\mathbb{C}^*}\), where
\begin{equation}
	\xi^{{\rm Gr}\ b_1,\cdots,b_k}_{a_1,\cdots,a_k} = \sum_{j=1}^{k}z^{l_{a_j+1}}\delta_{a_1,b_1} \cdots \delta_{a_j+1,b_j} \cdots \delta_{r_k,l_k}, \nonumber
\end{equation}
if \(a_1 < n\) and
\small
\begin{equation}
	\xi^{{\rm Gr}\ b_1,\cdots,b_k}_{a_1,\cdots,a_k} = 
		(-1)^{k-1} z^{l_0}\delta_{a_2,b_1} \cdots \delta_{a_k,b_{k-1}} \delta_{0,b_k} + \sum_{j=2}^{k}z^{l_{a_j+1}}\delta_{a_1,b_1} \cdots \delta_{a_j+1,b_j} \cdots \delta_{a_k,b_k}, \nonumber
\end{equation}
\normalsize
if \(a_1 = n\), then \(\xi^{\rm Gr}\) is the holomorphic data of the tt*-equation (\ref{Gr}).

\begin{prop}\label{prop4.5}
	Let \(\phi^{\rm Gr} := \left({\rm det}_{a_i,b_j}\left(\phi_{a_i,b_j}\right)\right)_{\substack{n \ge a_1 > \cdots > a_k \ge 0\\ n \ge b_1 > \cdots > b_k \ge 0}}\) be a compound matrix of \(\phi\), then \(\phi^{\rm Gr}\) is a solution of \(d \phi^{\rm Gr} = \phi^{\rm Gr}\xi^{\rm Gr}\) such that \(\phi^{\rm Gr}\) admits an Iwasawa factorization
	\begin{equation}
		\phi^{\rm Gr} = F^{\rm Gr} \phi_+^{\rm Gr}, \nonumber
	\end{equation}
	where \(F^{\rm Gr} = \left({\rm det}_{a_i,b_j}\left(F_{a_i,b_j}\right)\right)_{\substack{n \ge a_1 > \cdots > a_k \ge 0 \\ n \ge b_1 > \cdots > b_k \ge 0}}\) satisfies
	\begin{equation}
		\overline{{\rm det}_{a_i,b_j}\left(F_{a_i,b_j}\right)} = {\rm det}_{a_i,b_j}\left(F_{n-a_i,n-b_j}\right), \nonumber
	\end{equation}
	and \(\phi_+^{\rm Gr} = \left({\rm det}_{a_i,b_j}\left(\phi_{+,a_i,b_j}\right)\right)_{\substack{n \ge a_1 > \cdots > a_k \ge 0 \\ n \ge b_1 > \cdots > b_k \ge 0}}\) satisfies
	\begin{equation}
		\left. {\rm det}_{a_i,b_j}\left(\phi_{+,a_i,b_j}\right)\right|_{\lambda=0} = e^{u_{a_1,\cdots,a_k}/2}\delta_{a_1,b_1} \cdots \delta_{a_k,b_k}, \nonumber
	\end{equation}
	for some smooth functions \(u_{a_1,\cdots,a_k}: \mathbb{C}^* \rightarrow \mathbb{R}\ (n \ge a_1 > \cdots > a_k \ge 0)\).
\end{prop}
\begin{proof}
	Let \(\{e_j\}_{j=0}^{n}\) be the standard basis of \(\mathbb{R}^{n+1}\), then
	\begin{equation}
		\left(\phi e_{r_1}\right) \wedge \cdots \wedge \left(\phi e_{r_k}\right) = \sum_{n \ge p_1 > \cdots > p_k \ge 0}{\rm det}_{p_a,r_b}\left(\phi_{p_a,r_b}\right) \cdot e_{p_1} \wedge \cdots \wedge e_{p_k}. \nonumber
	\end{equation}
	By differentiating both sides, we have
	\scriptsize
	\begin{align}
		&\sum_{n \ge p_1 > \cdots > p_k \ge 0}\left(\frac{d}{dz}{\rm det}_{p_a,r_b}\left(\phi_{p_a,r_b}\right)\right) \cdot e_{p_1} \wedge \cdots \wedge e_{p_k} \nonumber\\
		&= \sum_{j=1}^{k}\left(\phi e_{r_1}\right) \wedge \cdots \left(\frac{d\phi}{dz}e_{r_j}\right) \wedge \cdots \wedge \left(\phi e_{r_k}\right) \nonumber\\
		&= \lambda^{-1}\sum_{j=1}^{k}z^{l_{r_j+1}}\left(\phi e_{r_1}\right) \wedge \cdots \left(\phi e_{r_j + 1}\right) \wedge \cdots \wedge \left(\phi e_{r_k}\right) \nonumber\\
		&= \lambda^{-1}\sum_{n \ge p_1 > \cdots > p_k \ge 0} \sum_{j=1}^{k}z^{l_{r_j+1}}{\rm det}\left(\phi_{p_a,r_b}\right)_{\substack{n \ge r_1 > \cdots > r_{j-1} \ge r_j+1 > r_{j+1} > \cdots > r_k \ge 0 \\ n \ge p_1 > \cdots > p_k \ge 0}} \cdot e_{p_1} \wedge \cdots \wedge e_{p_k}. \nonumber
	\end{align}
	\normalsize
	Thus, we have
	\footnotesize
	\begin{align}
		&\frac{d}{dz}{\rm det}_{p_a,r_b}\left(\phi_{p_a,r_b}\right) \nonumber\\
		&= \lambda^{-1}\sum_{j=1}^{k}z^{l_{r_j+1}}{\rm det}\left(\phi_{p_a,r_b}\right)_{\substack{n \ge r_1 > \cdots > r_{j-1} \ge r_j+1 > r_{j+1} > \cdots > r_k \ge 0 \\ n \ge p_1 > \cdots > p_k \ge 0}} \nonumber\\
		&= \sum_{n \ge q_1 > \cdots > q_k \ge 0} {\rm det}_{p_a,q_c}\left(\phi_{p_a,q_c}\right)_{\substack{n \ge q_1 > \cdots > q_k \ge 0 \\ n \ge p_1 > \cdots > p_k \ge 0}} \lambda^{-1}\sum_{j=1}^{k}z^{l_{r_j + 1}}\delta_{q_1,r_1} \cdots \delta_{q_j,r_j+1} \cdots \delta_{q_k,r_k}, \nonumber
	\end{align}
	\normalsize
	and then, we obtain \(d\phi^{\rm Gr} = \phi^{\rm Gr}\xi^{\rm Gr}\). Since \(\phi = F\phi_+\), we obtain \(\phi^{\rm Gr} = F^{\rm Gr}\phi_+^{\rm Gr}\) and \(u_{r_1,\cdots,r_k} = u_{r_1} + \cdots + u_{r_k}\), where \(\phi_+|_{\lambda = 0} = {\rm diag}(e^{u_1/2},\cdots,e^{u_n/2})\).
\end{proof}\vskip\baselineskip

As in Example \ref{ex2tt}, put
\scriptsize
\begin{align}
	&w_{r_1,\cdots,r_k} \nonumber\\
	&= u_{r_1,\cdots,r_k} - \frac{1}{n+1}\left\{-2(n+1)\sum_{j=1}^{k}\sum_{a=1}^{r_j}l_a + \left(2\sum_{j=1}^{k}r_j+1\right)\sum_{b=1}^{n}l_b + \left(2\sum_{j=1}^{k}r_j-n\right)l_0\right\}\log{|z|}, \nonumber
\end{align}
\normalsize
and \( t = \frac{n+1}{n+1+\sum_{a=0}^{n}l_a} z^{\frac{n+1+\sum_{a=0}^{n}l_a}{n+1}}\), then we have \(w_{r_1,\cdots,r_k} = w_{r_1} + \cdots + w_{r_k}\), where \(\{w_j\}_{j=0}^{n}\) is a solution to the tt*-Toda equation in Example \ref{ex2tt}. Thus, 
we obtain the following corollary.

\begin{cor}
	The \(\xi^{\rm Gr}\) is the holomorphic data of global radial solution \(\{w_{r_1,\cdots,r_k}\}_{n \ge r_ > \cdots > r_k \ge 0}\) to the tt*-equation (\ref{Gr}) with the condition (I), (II) and the asymptotic data
	\scriptsize
	\begin{equation}
		w_{r_1,\cdots,r_k} \sim - \frac{1}{n+1}\left\{-2(n+1)\sum_{j=1}^{k}\sum_{a=1}^{r_j}l_a + \left(2\sum_{j=1}^{k}r_j+1\right)\sum_{b=1}^{n}l_b + \left(2\sum_{j=1}^{k}r_j-n\right)l_0\right\}\log{|z|}, \nonumber
	\end{equation}
	\normalsize
	as \(z \rightarrow 0\).
\end{cor}
\begin{proof}
	Since \(\{w_j\}_{j=0}^{n}\) is a solution to the tt*-Toda equation, \(\{w_{r_1,\cdots,r_k}\}_{n \ge r_1 > \cdots > r_k \ge 0}\) is a solution to the tt*-equation (\ref{Gr}). Thus, the \(\xi^{\rm Gr}\) gives a solution to the tt*-equation (\ref{Gr}). Conversely, given a global radial solution to the tt*-equation (\ref{Gr}) with the asymptotic data \(\{m_{r_1,\cdots,r_k}\}_{n \ge r_1 > \cdots > r_k \ge 0}\). Put \(l_j = \frac{1}{2}(m^{j-1,n+1-j} - m^{j,n-j})\), then it follows from Proposition \ref{prop4.5} that \(\xi^{\rm Gr}\) gives the solutions.
\end{proof}\vskip\baselineskip

\begin{ex}[The holomorphic data of the tt*-equation (\ref{Gr}) for the case \(N = 2, k=2\)]\label{ex5}
	Let
	\begin{equation}
		\xi^{\rm Gr} = \frac{1}{\lambda}\left(\begin{array}{cccccc}
			0 & 0 & 0 & 0 & -z^{l_0} & 0 \\
			z^{l_2} & 0 & 0 & 0 & 0 & -z^{l_0} \\
			0 & z^{l_1} & 0 & 0 & 0 & 0 \\
			0 & z^{l_3} & 0 & 0 & 0 & 0 \\
			0 & 0 & z^{l_3} & z^{l_1} & 0 & 0 \\
			0 & 0 & 0 & 0 & z^{l_2} & 0
		\end{array}\right)\frac{dz}{z}, \nonumber
	\end{equation}
	where \(l_j \ge -1\), \(l_1 = l_3\) and \(l_0 + l_1 + l_2 + l_3 > -4\). The \(\xi^{\rm Gr}\) is the holomorphic data of the solution in Example \ref{ex4}.
	\qed
\end{ex}\vskip\baselineskip

Hence, \(\xi^{\rm Gr}\) gives the tt*-equation (\ref{Gr}) by using the DPW method. This DPW potential \(\xi^{\wedge^k}\) describes the tt*-structure on \(qH^*({\rm Gr}(k,\mathbb{C}^{k+N}))\).

\section{The relation between \(qH^*({\rm Gr}(k,\mathbb{C}^{k+N})) and \bigwedge^kqH^*(\mathbb{C}P^n)\)}
In \cite{B1995}, Bourdeau described a relation between the tt*-equation for \(qH^*({\rm Gr}(k,\mathbb{C}^{k+N}))\) and the tt*-equation for \(qH^*(\mathbb{C}P^{k+N-1})\). In this section, we give a tt*-structure on the exterior product od \(E^{\mathbb{C}P}_n\) and we show that the tt*-structure on \(\bigwedge^k E^{\mathbb{C}P}_n\) is isomorphic to the tt*-structure on \(\bigwedge^k E^{\mathbb{C}P}_n\) as tt*-structures. In section 5.1, we show that the \(k\)-th exterior product of the tt*-structure \((E^{\mathbb{C}P}_n,\eta^{\mathbb{C}P},g^{\mathbb{C}P},\Phi^{\mathbb{C}P})\) is a tt*-structure. In section 5.2, we give an isomorphism of tt*-structure between \((E^{\rm Gr},\eta^{\rm Gr},g^{\rm Gr},\Phi^{\rm Gr})\) and \((\bigwedge^k E^{\mathbb{C}P}_n,\eta^{\wedge^k},g^{\wedge^k},\Phi^{\wedge^k})\). In section 5.3, we give a Lie-theoretic description of the tt*-structures.

\subsection{The induce tt*-structure on \(\bigwedge^k E^{\mathbb{C}P}_n\)}
Let \((E^{\mathbb{C}P}_n,\eta^{\mathbb{C}P},g^{\mathbb{C}P},\Phi^{\mathbb{C}P})\) be the tt*-structure in section 3.1, then we define a bilinear form \(\eta^{\wedge^k}\), a Hermitian form \(g^{\wedge^k}\) on \(\bigwedge^k E_n^{\mathbb{C}P}\) and an \({\rm End}(\bigwedge^k E_n^{\mathbb{C}P})\)-valued 1-form by
\begin{align}
	&\eta^{\wedge^k}(a_1 \wedge \cdots \wedge a_k,b_1 \wedge \cdots \wedge b_k) = {\rm det}\left(\eta^{\mathbb{C}P}(a_i,b_j)\right), \nonumber\\
	&g^{\wedge^k}(a_1 \wedge \cdots \wedge a_k,b_1 \wedge \cdots \wedge b_k) = {\rm det}\left(g^{\mathbb{C}P}(a_i,b_j)\right), \nonumber\\
	&\Phi^{\wedge^k}(a_1 \wedge \cdots \wedge a_k) = \sum_{j=1}^{k}a_1 \wedge \cdots \wedge \left(\Phi^{\mathbb{C}P}(a_j) \right) \wedge \cdots \wedge a_k. \nonumber
\end{align}
We have the following lemma.

\begin{lem}
	We have
	\begin{itemize}
		\item [(1)] \(\eta^{\wedge^k}\) is nondegenerate,
		
		\item [(2)] \(g^{\wedge^k}\) is positive-definite. 
	\end{itemize}
\end{lem}
\begin{proof}
	Since \({\rm det}\left(\eta^{\mathbb{C}P}(a_i,b_j)\right)\) is the \(k\)-th compound matrix of \(\left(\eta^{\mathbb{C}P}(a_i,b_j)\right)_{0 \le i,j \le n}\) and \(\eta^{\mathbb{C}P}\) is nondegenerate, we obtain (1). Since \({\rm det}\left(g^{\mathbb{C}P}(a_i,b_j)\right)\) is the \(k\)-th compound matrix of \(\left(g^{\mathbb{C}P}(a_i,b_j)\right)_{0 \le i,j \le n}\) and \(g^{\mathbb{C}P}\) is positive-definite, we obtain (2).
\end{proof}

We show that \((\bigwedge^k E^{\mathbb{C}P}_n,\eta^{\wedge^k},g^{\wedge^k},\Phi^{\wedge^k})\) is a tt*-structure. We denote the real form and the holomorphic structure of \((E^{\mathbb{C}P}_n,\eta^{\mathbb{C}P},g^{\mathbb{C}P},\Phi^{\mathbb{C}P})\) by \(\kappa^{\mathbb{C}P}\) and \(\overline{\partial}_{E^{\mathbb{C}P}_n}\) respectively. 
\begin{lem}
	We have
	\begin{itemize}
		\item [(i)] \(\Phi^{\wedge^k}\) is self-adjoint with respect to \(\eta^{\wedge^k}\),
		
		\item [(ii)] \(g^{\wedge^k}(a_1 \wedge \cdots \wedge a_k,b_1\wedge \cdots \wedge b_k) = \eta^{\wedge^k}(\kappa^{\mathbb{C}P}(a_1) \wedge \cdots \wedge \kappa^{\mathbb{C}P}(a_k),b_1 \wedge \cdots \wedge b_k)\),
		
		\item [(iii)] \(\left(\Phi^{\wedge^k}\right)^{\dagger_{g^{\wedge^k}}}(a_1 \wedge \cdots \wedge a_k) = \sum_{j=1}^{k}a_1 \wedge \cdots \wedge \left(\Phi^{\mathbb{C}P}\right)^{\dagger_{g^{\mathbb{C}P}}}(a_j) \wedge \cdots \wedge a_k\),
		
		\item [(iv)] {\small \(\partial\left(g^{\wedge^k}(a_1 \wedge \cdots \wedge a_k,b_1 \wedge \cdots \wedge b_k)\right) 
			= g^{\wedge^k}\left(a_1 \wedge \cdots \wedge a_k,\sum_{j=1}^{k} b_1 \wedge \cdots \wedge \left(\partial_{E^{\mathbb{C}P}_n}^{\dagger_{g^{\mathbb{C}P}}} b_j\right) \wedge \cdots \wedge b_k\right)\)},
	\end{itemize}
	for \(a_1,\cdots,a_k,b_1,\cdots,b_k \in 
	\Gamma(E^{\mathbb{C}P}_n)\).
\end{lem}
\begin{proof}
	(i) We have
	\begin{align}
		&\eta^{\wedge^k}\left(\Phi^{\wedge^k}(a_1 \wedge \cdots \wedge a_k),b_1 \wedge \cdots \wedge b_k\right) \nonumber\\
		&= \sum_{j=1}^{k} \sum_{\sigma \in \mathfrak{S}_k} {\rm sgn}(\sigma) \eta(a_1,b_{\sigma(1)}) \cdots \eta\left(\Phi^{\mathbb{C}P}(a_j),b_{\sigma(j)}\right) \cdots \eta(a_k,b_{\sigma(k)}) \nonumber\\
		&= \sum_{j=1}^{k} \sum_{\sigma \in \mathfrak{S}_k} {\rm sgn}(\sigma) \eta(a_1,b_{\sigma(1)}) \cdots \eta\left(a_j,\Phi^{\mathbb{C}P}(b_{\sigma(j)})\right) \cdots \eta(a_k,b_{\sigma(k)}) \nonumber\\
		&= \eta^{\wedge^k}\left(a_1 \wedge \cdots \wedge a_k,\Phi^{\wedge^k}(b_1 \wedge \cdots \wedge b_k)\right) \nonumber
	\end{align}\vspace{2mm}
	
	(ii) From \(g^{\mathbb{C}P}(a,b) = \eta^{\mathbb{C}P}(\kappa^{\mathbb{C}P}(a),b)\), we have
	\begin{align}
		g^{\wedge^k}(a_1 \wedge \cdots \wedge a_k,b_1 \wedge \cdots \wedge b_k) &= {\rm det}(g^{\mathbb{C}P}(a_i,b_j)) = {\rm det}(\eta^{\mathbb{C}P}(\kappa(a_i),b_j)) \nonumber\\
		&= \eta^{\wedge^k}\left(\kappa^{\mathbb{C}P}(a_1) \wedge \cdots \wedge \kappa^{\mathbb{C}P}(a_k),b_1 \wedge \cdots \wedge b_k\right). \nonumber
	\end{align}
	
	(iii) Since
	\footnotesize
	\begin{align}
		&g^{\wedge^k}\left(a_1 \wedge \cdots \wedge a_k,\left(\Phi^{\wedge^k}\right)^{\dagger_{g^{\wedge^k}}}(b_1 \wedge \cdots \wedge b_k)\right) 
		=g^{\wedge^k}\left(\Phi^{\wedge^k}(a_1 \wedge \cdots \wedge a_k),b_1 \wedge \cdots \wedge b_k\right) \nonumber\\
		&= \sum_{j=1}^{k} \sum_{\sigma \in \mathfrak{S}_k} {\rm sgn}(\sigma) g^{\mathbb{C}P}(a_1,b_{\sigma(1)}) \cdots g^{\mathbb{C}P}\left(\Phi^{\mathbb{C}P}(a_j),b_{\sigma(j)}\right) \cdots g^{\mathbb{C}P}(a_k,b_{\sigma(k)}) \nonumber\\
		&= \sum_{j=1}^{k} \sum_{\sigma \in \mathfrak{S}_k} {\rm sgn}(\sigma) g^{\mathbb{C}P}(a_1,b_{\sigma(1)}) \cdots g^{\mathbb{C}P}\left(a_j,\left(\Phi^{\mathbb{C}P}\right)^{\dagger_{g^{\mathbb{C}P}}}(b_{\sigma(j)})\right) \cdots g^{\mathbb{C}P}(a_k,b_{\sigma(k)}) \nonumber\\
		&= g^{\wedge^k}\left(a_1 \wedge \cdots \wedge a_k,\sum_{j=1}^{k}b_1 \wedge \cdots \wedge \left(\Phi^{\mathbb{C}P}\right)^{\dagger_{g^{\mathbb{C}P}}}(b_j) \wedge \cdots \wedge b_k\right), \nonumber
	\end{align}
	\normalsize
	we obtain
	\begin{equation}
		\left(\Phi^{\wedge^k}\right)^{\dagger_{g^{\wedge^k}}}(b_1 \wedge \cdots \wedge b_k) = \sum_{j=1}^{k}b_1 \wedge \cdots \wedge \left(\Phi^{\mathbb{C}P}\right)^{\dagger_{g^{\mathbb{C}P}}}(b_j) \wedge \cdots \wedge b_k. \nonumber
	\end{equation}\vspace{2mm}
	
	(iv) We have
	\begin{align}
		&\partial\left(g^{\wedge^k}(a_1 \wedge \cdots \wedge a_k,b_1 \wedge \cdots \wedge b_k)\right) \nonumber\\
		&= \partial \left(\sum_{\sigma \in \mathfrak{S}_k} {\rm sgn}(\sigma) g^{\mathbb{C}P}(a_1,b_{\sigma (1)}) \cdots g^{\mathbb{C}P}(a_k,b_{\sigma (k)})\right) \nonumber\\
		&= \sum_{j=1}^{k} \sum_{\sigma \in \mathfrak{S}_k} {\rm sgn}(\sigma) g^{\mathbb{C}P}(a_1,b_{\sigma (1)}) \cdots g^{\mathbb{C}P}\left(a_j,\partial_{E^{\mathbb{C}P}_n }^{\dagger_{g^{\mathbb{C}P}}}b_{\sigma}\right) \cdots g^{\mathbb{C}P}(a_k,b_{\sigma (k)}) \nonumber\\
		&= g^{\wedge^k}\left(a_1 \wedge \cdots \wedge a_k,\sum_{j=1}^{k} b_1 \wedge \cdots \wedge \left(\partial_{E^{\mathbb{C}P}_n}^{\dagger_{g^{\mathbb{C}P}}} b_j\right) \wedge \cdots \wedge b_k\right). \nonumber
	\end{align}
\end{proof}
We obtain the following theorem.
\begin{thm}\label{thm4.1}
	\((\bigwedge^k E^{\mathbb{C}P}_n,\eta^{\wedge^k},g^{\wedge^k},\Phi^{\wedge^k})\) is a tt*-structure.
\end{thm}
\begin{proof}
	From Lemma \ref{lem4.1}, we obtain (a), (b) in Definition 2.1. We show the condition (c). Let \(\overline{\partial}_{E^{\wedge^k}}\) be the holomorphic structure on \(\bigwedge^{k} E^{\mathbb{C}P}_n\). We put
	\begin{equation}
		\nabla^{\wedge^k} = \partial_{E^{\wedge^k}}^{\dagger_{g^{\wedge^k}}} + \overline{\partial}_{E^{\wedge^k}} + \lambda^{-1} \Phi^{\wedge^k} + \lambda \left(\Phi^{\wedge^k}\right)^{\dagger_{g^{\wedge^k}}}, \nonumber
	\end{equation}
	then from Lemma \ref{lem4.1} we have
	\begin{equation}
		\nabla^{\wedge^k}(a_1 \wedge \cdots \wedge a_k) = \sum_{j=1}^{k} a_1 \wedge \cdots \wedge \nabla^{\mathbb{C}P} a_j \wedge \cdots \wedge a_k, \nonumber
	\end{equation}
	where \(\nabla^{\mathbb{C}P} = \partial_{E^{\mathbb{C}P}_n}^{\dagger_{g^{\mathbb{C}P}}} + \overline{\partial}_{E^{\mathbb{C}P}_n} + \lambda^{-1} \Phi^{\mathbb{C}P} + \lambda \left(\Phi^{\mathbb{C}P}\right)^{\dagger_{g^{\mathbb{C}P}}}\). Put \(\nabla^{\mathbb{C}P}e_j = \sum_{l=1}^{k}\alpha_{lj}e_l\), then we obtain
	\footnotesize
	\begin{align}
		&\left(\nabla^{\wedge^k}\right)^2(e_{l_1} \wedge \cdots \wedge e_{l_k}) 
		= \sum_{i,j=1}^{k} \sum_{a,b=1}^{k} \alpha_{a,l_j} a_{b,l_i} e_{l_1} \wedge \cdots \wedge \overbrace{e_a}^{j} \wedge \cdots \wedge \overbrace{e_b}^{i} \wedge \cdots \wedge e_{l_k} \nonumber\\
		&= \sum_{1 \ge i > j \ge k} \sum_{a,b=1}^{k} \alpha_{a,l_j} \wedge \alpha_{b,l_i} \cdot e_{l_1} \wedge \cdots \wedge \overbrace{e_a}^{j} \wedge \cdots \wedge \overbrace{e_b}^{i} \wedge \cdots \wedge e_{l_k} \nonumber\\
		&\ \ \ \ \ \sum_{1 \le i < j \le k} \sum_{a,b=1}^{k} \alpha_{a,l_i} \wedge \alpha_{b,l_j} \cdot e_{l_1} \wedge \cdots \wedge \overbrace{e_b}^{i} \wedge \cdots \wedge \overbrace{e_a}^{j} \wedge \cdots \wedge e_{l_k} \nonumber\\&= 0. \nonumber
	\end{align}
	\normalsize
	Hence, \(\nabla^{\wedge^k}\) is flat.
\end{proof}\vskip\baselineskip

\begin{ex}[The quantum cohomology of \(\bigwedge^2 \mathbb{C}P^3\)]\label{ex3}
	We consider
	\begin{equation}
		\bigwedge^2 E^{\mathbb{C}P}_3 = \bigsqcup_{z \in \mathbb{C}^*}\bigwedge^2 \mathbb{C}[X]/<X^4-z>, \nonumber
	\end{equation}
	with the frame \(e = (e_1 \wedge e_0,e_2 \wedge e_0,e_2 \wedge e_1,e_3 \wedge e_0,e_3 \wedge e_1,e_3 \wedge e_2)\), where
	\begin{equation}
		(e_{r_1} \wedge e_{r_2})(z) = (z,[X^{r_1} \wedge X^{r_2}]),\ \ \ 3 \ge r_1 > r_2 \ge 0.\nonumber
	\end{equation}
	Then the non-vanishing elements of \(\eta^{\wedge^2}(e_{l_1} \wedge e_{l_2},e_{r_1} \wedge e_{r_2})\) are
	\begin{align}
		\eta^{\rm \wedge^2}(e_1 \wedge e_0,e_3 \wedge e_2) &= \eta^{\rm \wedge^2}(e_2 \wedge e_0,e_3 \wedge e_1) = \eta^{\rm \wedge^2}(e_2 \wedge e_1,e_2 \wedge e_1) \nonumber\\
		&= \eta^{\rm \wedge^2}(e_3 \wedge e_0,e_3 \wedge e_0) = \eta^{\rm \wedge^2}(e_3 \wedge e_1,e_2 \wedge e_0) \nonumber\\
		&= \eta^{\rm \wedge^2}(e_3 \wedge e_2,e_1 \wedge e_0) \nonumber\\ 
		&= -1, \nonumber
	\end{align}
	\(g^{\wedge^2}(e_{r_1} \wedge e_{r_2},e_{l_1} \wedge e_{l_2}) = e^{w_{r_1} + w_{r_2}}\delta_{r_1,l_1}\delta_{r_2,l_2}\),
	\begin{align}
		&\Phi^{\wedge^2}e = e \cdot \left(\begin{array}{cccccc}
			0 & 0 & 0 & 0 & z & 0 \\
			1 & 0 & 0 & 0 & 0 & z \\
			0 & 1 & 0 & 0 & 0 & 0 \\
			0 & 1 & 0 & 0 & 0 & 0 \\
			0 & 0 & 1 & 1 & 0 & 0 \\
			0 & 0 & 0 & 0 & 1 & 0
		\end{array}\right)\frac{dz}{z}. \nonumber
	\end{align}
	and 
	\footnotesize
	\begin{align}
		&\Phi^{\dagger_{g^{\rm Gr}}}(e) \nonumber\\
		&= e \cdot \left(\begin{array}{cccccc}
			0 & e^{u_{2,0} - u_{1,0}} & 0 & 0 & 0 & 0 \\
			0 & 0 & e^{u_1 - u_0} & e^{u_3 - u_2} & 0 & 0 \\
			0 & 0 & 0 & 0 & e^{u_3- u_2} & 0 \\
			0 & 0 & 0 & 0 & e^{u_1 - u_0} & 0 \\
			\overline{z}e^{u_0- u_3} & 0 & 0 & 0 & 0 & e^{u_2- u_1}\\
			0 & \overline{z}e^{u_0 - u_3} & 0 & 0 & 0 & 0
		\end{array}\right)\frac{d\overline{z}}{\overline{z}}, \nonumber
	\end{align}
	\normalsize
	where \(\{u_j\}_{j=0}^{3}\) satisfies
	\begin{equation}
		\left\{
		\begin{array}{l}
			(u_0)_{z\overline{z}} = e^{u_0-u_3} - |z|^{-2}e^{u_1-u_0},\\
			(u_1)_{z\overline{z}} = |z|^{-2}e^{u_1-u_0} - |z|^{-2}e^{u_2-u_1},\\
			(u_2)_{z\overline{z}} = |z|^{-2}e^{u_2-u_1} - |z|^{-2}e^{u_3-u_1},\\
			(u_3)_{z\overline{z}} = |z|^{-2}e^{u_3-u_2} - e^{u_0-u_3},
		\end{array}
		\right. \nonumber
	\end{equation}
	with the condition \(u_3=-u_0, u_3=-u_2\). Then \((\bigwedge^2 E^{\mathbb{C}P}_3,\eta^{\wedge^2},g^{\wedge^2},\Phi^{\wedge^2})\) is a tt*-structure.
	\qed
\end{ex}\vskip\baselineskip

Thus, we obtain a tt*-structure on \(\bigwedge^k E^{\mathbb{C}P}_n\) from the tt*-structure on \(E^{\mathbb{C}P}_n\). In the following section, we show that the tt*-structure on \(\bigwedge^k E^{\mathbb{C}P}_n\) is isomorphic to the tt*-structure on \(E^{\rm Gr}\) in section 4.

\subsection{The relation between the tt*-structure on \(E^{\rm Gr}_{k,N}\) and the tt*-structure on \(\bigwedge^k E^{\rm Gr}_n\)}
In this section, we prove that the tt*-structure on \(E^{\rm Gr}_{k,N}\) isomorphic to the tt*-structure on \(\bigwedge^k E^{\rm Gr}_n\) as tt*-structures by an isomorphism
\begin{equation}
	\mathcal{T}^{\wedge^k}: \bigwedge^k E^{\mathbb{C}P}_n \stackrel{\sim}{\longrightarrow} E^{\rm Gr}_{k,N}: e_{r_1} \wedge \cdots \wedge e_{r_k} \mapsto e_{r_1,\cdots,r_k}. \nonumber
\end{equation}

First, we prove (1), (2) of Definition \ref{def2.2}.
\begin{lem}\label{lem4.2}
	We have
	\begin{align}
		&\eta^{\wedge^k}(a_1 \wedge \cdots \wedge a_k,b_1 \wedge \cdots \wedge b_k) = \eta^{\rm Gr}(\mathcal{T}^{\wedge^k}(a_1 \wedge \cdots \wedge a_k),\mathcal{T}^{\wedge^k}(b_1 \wedge \cdots \wedge b_k)), \nonumber\\
		&g^{\wedge^k}(a_1 \wedge \cdots \wedge a_k,b_1 \wedge \cdots \wedge b_k) = g^{\rm Gr}(\mathcal{T}^{\wedge^k}(a_1 \wedge \cdots \wedge a_k),\mathcal{T}^{\wedge^k}(b_1 \wedge \cdots \wedge b_k)), \nonumber
	\end{align}
	for all \(a_1,\cdots,a_k, b_1,\cdots,b_k \in \Gamma(E^{\mathbb{C}P}_n)\).
\end{lem}
\begin{proof}
	From the definition, we have
	\begin{align}
		&\eta^{\wedge^k}(e_{l_1} \wedge \cdots \wedge e_{l_k},e_{r_1} \wedge \cdots \wedge e_{r_k}) \nonumber\\
		&= {\rm det}\left(\eta^{\mathbb{C}P}(e_{l_i},e_{r_j})\right) = \sum_{\sigma \in \mathfrak{S}_k} {\rm sgn}(\sigma) \eta^{\mathbb{C}P}(e_{l_1},e_{r_\sigma(1)}) \cdots \eta^{\mathbb{C}P}(e_{l_k},e_{r_{\sigma(k)}}) \nonumber\\
		&= \sum_{\sigma \in \mathfrak{S}_k} {\rm sgn}(\sigma)\delta_{l_1,n-r_{\sigma(1)}} \cdots \delta_{l_k,n-r_{\sigma(n)}} = (-1)^{\left[\frac{k}{2}\right]}\delta_{l_1,n-r_k} \cdots \delta_{l_k,n-r_1} \nonumber\\
		&=\eta^{\rm Gr}(\mathcal{T}^{\wedge^k}(e_{l_1} \wedge \cdots \wedge e_{l_k}),\mathcal{T}^{\wedge^k}(e_{r_1} \wedge \cdots \wedge e_{r_k})), \nonumber
	\end{align}
	and
	\footnotesize
	\begin{align}
		g^{\wedge^k}(e_{r_1} \wedge \cdots \wedge e_{r_k},e_{l_1} \wedge \cdots \wedge e_{l_k}) &= {\rm det}\left(g^{\mathbb{C}P}(e_{r_i},e_{l_j})\right) = {\rm det}\left(e^{w_{r_i}}\delta_{r_i,l_j}\right) \nonumber\\
		&= e^{w_{r_1} + \cdots + w_{r_k}}\delta_{r_1,n-l_k} \cdots \delta_{r_k,n-l_1} \nonumber\\
		&= g^{\rm Gr}(T^{\wedge^k}(e_{r_1} \wedge \cdots \wedge e_{r_k}),T^{\wedge^k}(e_{l_1} \wedge \cdots \wedge e_{l_k})). \nonumber
	\end{align}
	\normalsize
	Thus, we obtain the results stated above.
\end{proof}
Next, we show the condition (3) of Definition \ref{def2.2}.
\begin{lem}\label{lem4.3}
	We have
	\begin{equation}
		\mathcal{T}^{\wedge^k}(\Phi^{\wedge^k}_{z\frac{d}{dz}}(a_1 \wedge \cdots \wedge a_k)) = \Phi^{\rm Gr}_{z\frac{d}{dz}}(\mathcal{T}^{\wedge^k}(a_1 \wedge \cdots \wedge a_k)). \nonumber
	\end{equation}
\end{lem}
\begin{proof}
	Since \(s_{\mu_1,\cdots,\mu_k}\) is the Schur polynomial, we have
	\begin{align}
		X_1 \cdot s_{\mu_1,\cdots,\mu_k}&= (t_1 + \cdots + t_k) \cdot s_{\mu_1,\cdots,\mu_k} 
		= \sum_{j=1}^{k} s_{\mu_1,\cdots,\mu_j+1,\cdots,\mu_k}. \nonumber
	\end{align}
	When \(t_1^{n+1} = \cdots = t_k^{n+1} = (-1)^{k+1}z\), we have
	\begin{equation}
		s_{n-k+2,\mu_2,\cdots,\mu_k}(t_1,\cdots,t_k) = z s_{\mu_2-1,\cdots,\mu_k-1,0}(t_1,\cdots,t_k). \nonumber
	\end{equation}
	Then if \(r_1 < n\), we obtain
	\begin{align}
		&\mathcal{T}^{\wedge^k}\left(\Phi^{\wedge^k}_{z\frac{d}{dz}}(e_{r_1} \wedge \cdots \wedge e_{r_k})\right) \nonumber\\
		&= \sum_{j=1}^{k}e_{r_1,\cdots,r_j+1,\cdots,r_k} = X_1\cdot e_{r_1,\cdots,r_k} 
		= \Phi^{\rm Gr}_{z\frac{d}{dz}}(\mathcal{T}^{\wedge^k}(e_{r_1} \wedge \cdots \wedge e_{r_k})), \nonumber
	\end{align}
	and if \(r_1 = n\), we obtain
	\begin{align}
		\mathcal{T}^{\wedge^k}\left(\Phi^{\wedge^k}_{z\frac{d}{dz}}(e_n \wedge e_{r_2} \wedge \cdots \wedge e_{r_k})\right) &= ze_{r_2,\cdots,r_k,0} + \sum_{j=2}^{k}e_{n,r_2,\cdots,r_j+1,\cdots,r_k} \nonumber\\
		&= X_1\cdot e_{n,r_2,\cdots,r_k} \nonumber\\
		&= \Phi^{\rm Gr}_{z\frac{d}{dz}}(\mathcal{T}^{\wedge^k}(e_n\wedge e_{r_2} \wedge \cdots \wedge e_{r_k}). \nonumber
	\end{align}
\end{proof}
From Lemma \ref{lem4.2} and Lemma \ref{lem4.3}, we obtain isomorphism of tt*-structures between \\ \((E^{\rm Gr},\eta^{\rm Gr},g^{\rm Gr},\Phi^{\rm Gr})\) and \((\bigwedge^k E^{\mathbb{C}P}_n,\eta^{\wedge^k},g^{\wedge^k},\Phi^{\wedge^k})\).

\begin{thm}\label{thm4.2}
	Let \((E^{\rm \mathbb{C}P}_n,\eta^{\rm \mathbb{C}P},g^{\rm \mathbb{C}P},\Phi^{\rm \mathbb{C}P})\) be the tt*-structure in section 3.1 with the holomorphic data
	\begin{equation}
		\xi = \frac{1}{\lambda}\left(\begin{array}{cccc}
			& & & z^{l_0} \\
			z^{l_1} & & \\
			& \ddots & & \\
			& & z^{l_n} &
		\end{array}\right)dz, \nonumber
	\end{equation}
	and \((E^{\rm Gr}_{k,N},\eta^{\rm Gr},g^{\rm Gr},\Phi^{\rm Gr})\) the tt*-structure in section 3.2 with the holomorphic data \(\xi^{\rm Gr} = \left(\lambda^{-1}\xi^{{\rm Gr}\ b_1,\cdots,b_k}_{a_1,\cdots,a_k}\right)_{\substack{n \ge r_1 > \cdots > r_k \ge 0 \\ n \ge b_1 > \cdots > b_k \ge 0}}dz \in (\Lambda {\mathfrak{sl}}_{\binom{n}{k}} \mathbb{C})_{\sigma} \otimes \Omega^{1,0}_{\mathbb{C}^*}\), where
	\begin{equation}
		\xi^{{\rm Gr}\ b_1,\cdots,b_k}_{a_1,\cdots,a_k} = \sum_{j=1}^{k}z^{l_{a_j+1}}\delta_{a_1,b_1} \cdots \delta_{a_j+1,b_j} \cdots \delta_{r_k,l_k}, \nonumber
	\end{equation}
	if \(a_1 < n\) and
	\small
	\begin{equation}
		\xi^{{\rm Gr}\ b_1,\cdots,b_k}_{a_1,\cdots,a_k} = 
		(-1)^{k-1} z^{l_0}\delta_{a_2,b_1} \cdots \delta_{a_k,b_{k-1}} \delta_{0,b_k} + \sum_{j=2}^{k}z^{l_{a_j+1}}\delta_{a_1,b_1} \cdots \delta_{a_j+1,b_j} \cdots \delta_{a_k,b_k}, \nonumber
	\end{equation}
	\normalsize
	if \(a_1 = n\). Then, \(T^{\wedge^k}\) is an isomorphism of tt*-structures
	\begin{align}
		&\mathcal{T}^{\wedge^k}: \left(\bigwedge^k E^{\mathbb{C}P}_n,\eta^{\wedge^k},g^{\wedge^k},\Phi^{\wedge^k}\right) \stackrel{\sim}{\longrightarrow} (E^{\rm Gr}_{k,N},\eta^{\rm Gr},g^{\rm Gr},\Phi^{\rm Gr}). \nonumber		
	\end{align}
\end{thm}
\begin{proof}
	It follows from Lemma \ref{lem4.2} and Lemma \ref{lem4.3}.
\end{proof}\vskip\baselineskip

From the viewpoint of physics \cite{B1995}, this tt*-equation describe the ground state metric for the Grassmannian \(\sigma\)-model.\vskip\baselineskip
\begin{ex}[The tt*-structure on \(qH^*({\rm Gr}(2,\mathbb{C}^4))\)]
	Let \((E^{\rm Gr}_{2,2},\eta^{\rm Gr},g^{\rm Gr},\Phi^{\rm Gr})\) the tt*-structure in Example \ref{ex2} and \((\bigwedge^2 E^{\mathbb{C}P}_3,\eta^{\wedge^2},g^{\wedge^2},\Phi^{\wedge^2})\) be the tt*-structure in Example \ref{ex3}. We consider a bundle map
	\begin{equation}
		\mathcal{T}^{\wedge^2}: \bigwedge^2 E^{\mathbb{C}P}_3 \longrightarrow E^{\rm Gr}_{2,2}: e_{r_1} \wedge e_{r_2} \mapsto e_{r_1,r_2}. \nonumber
	\end{equation}
	Then, \((E^{\rm Gr},\eta^{\rm Gr},g^{\rm Gr},\Phi^{\rm Gr})\) is an isomorphism of tt*-structures
	\begin{equation}
		\mathcal{T}^{\wedge^2}: (\bigwedge^2 E^{\mathbb{C}P}_3,\eta^{\wedge^2},g^{\wedge^2},\Phi^{\wedge^2}) \longrightarrow (E^{\rm Gr}_{2,2},\eta^{\rm Gr},g^{\rm Gr},\Phi^{\rm Gr}). \nonumber
	\end{equation}
	\qed
\end{ex}

\subsection{A tt*-structure on a principal \(G\)-bundle}
We consider a principal \(G\)-bundle constructed from the tt*-Toda equation. In this section, we explain the relation between our construction and the result of Guest \cite{G2021}. We review a Lie theoretic description introduced by Guest, Lin \cite{GL2012}, \cite{GL2014} in terms of principal \(G\)-bundles.\vskip\baselineskip

Let \(G^{\mathbb{C}}= {\rm GL}_{n+1}\mathbb{C}\) and \(\mathfrak{g}^{\mathbb{C}}=\mathfrak{gl}_{n+1}\mathbb{C}\). We consider a complex-linear involution \(\sigma:\mathfrak{g}^{\mathbb{C}} \rightarrow \mathfrak{g}^{\mathbb{C}}\)
\begin{equation}
	\sigma(X) = -\Delta^{-1} X^t \Delta,\ \ \ X \in \mathfrak{g}^{\mathbb{C}}, \nonumber
\end{equation}
and a complex conjugate-linear involution \(c:\mathfrak{g}^{\mathbb{C}} \rightarrow \mathfrak{g}^{\mathbb{C}}\)
\begin{equation}
	c(X) = S \overline{X} \overline{S},\ \ \ X \in \mathfrak{g}^{\mathbb{C}},\ \ \ \Delta = S = \left(\begin{array}{ccc}
		& & 1 \\
		& \iddots & \\
		1 & &
	\end{array}\right). \nonumber \nonumber
\end{equation}
We put
\begin{align}
	&\mathfrak{k}^{\mathbb{C}} = \{X \in \mathfrak{g}^{\mathbb{C}}\ |\ \sigma(X) = X\},\ \ \ \ \ \mathfrak{p}^{\mathbb{C}} = \{X \in \mathfrak{g}^{\mathbb{C}}\ | \ \sigma(X) = -X\},\nonumber
\end{align}
then, we define a principal \(G^{\mathbb{C}}\)-bundle \(P = \mathbb{C}^* \times G^{\mathbb{C}}\), a \(\mathfrak{p}^{\mathbb{C}}\)-valued 1-form \(\Phi^P \in \Omega^1(\mathbb{C}^*,\mathfrak{p}^{\mathbb{C}})\)
\begin{equation}
	\Phi^P = \left(\begin{array}{ccc}
		e^{\frac{u_0}{2}} & & \\
		& \ddots & \\
		& & e^{\frac{u_n}{2}}
	\end{array}\right)\left(\begin{array}{cccc}
		& & & z\\
		1 & & \\
		& \ddots & & \\
		& & 1 &
	\end{array}\right)\left(\begin{array}{ccc}
		e^{-\frac{u_0}{2}} & & \\
		& \ddots & \\
		& & e^{-\frac{u_n}{2}}
	\end{array}\right)\frac{dz}{z}, \nonumber
\end{equation}
and a connection \(A^P \in \Omega^1(\mathbb{C}^*,\mathfrak{k}^{\mathbb{C}})\)
\begin{equation}
	A^P = \left(\begin{array}{ccc}
		\frac{1}{2}(u_0)_z& & \\
		& \ddots & \\
		& & \frac{1}{2}(u_n)_z
	\end{array}\right)dz - \left(\begin{array}{ccc}
		\frac{1}{2}(u_0)_{\overline{z}}& & \\
		& \ddots & \\
		& & \frac{1}{2}(u_n)_{\overline{z}}
	\end{array}\right)d\overline{z}, \nonumber
\end{equation}
for \(z \in \mathbb{C}^*\) and some functions \(u_j:\mathbb{C}^* \rightarrow \mathbb{R}\ (j=0,\cdots,n)\) with \(u_j + u_{n-j} = 0\). The Hitchin's equation for \((A^P,\Phi^P)\) is equivalent to the tt*-Toda equation.

\begin{prop}
	The pair \((A^P,\Phi^P)\) satisfies the Hitchin's equation
	\begin{equation}
		\left\{
		\begin{array}{l}
			F_{A^P} + [\Phi^P,c(\Phi^P)] = 0,\\
			\overline{\partial}_{A^P}\Phi^P = 0,
		\end{array}
		\right. \nonumber
	\end{equation}
	where \(F_{A^P}\) is the curvature form of \(A^P\), if and only if \(\{u_j\}_{j=0}^{n}\) is a solution of the tt*-Toda equation
	\begin{equation}
		\left\{\begin{array}{ll}
			(u_0)_{z\overline{z}} = e^{u_0-u_k} - |z|^{-2}e^{u_1-u_0}, & \\
			(u_j)_{z\overline{z}} = |z|^{-2}e^{u_j-u_{j-1}} - |z|^{-2}e^{u_{j+1}-u_j}, & j=1,\cdots,n,\\
			(u_k)_{z\overline{z}} = |z|^{-2}e^{u_k-u_{k-1}} - e^{u_0-u_k}.
		\end{array}
		\right. \nonumber
	\end{equation}
\end{prop}
\begin{proof}
	Direct calculation.
\end{proof}
Thus, \((P,c,\sigma,\Phi^P,A^P)\) describes the tt*-Toda equation as a principal bundle. The tt*-structure can be induced by \((P,c,\sigma,\Phi^P,A^P)\) as follows.\\

Let \(\rho:G^{\mathbb{C}} \rightarrow {\rm GL}_l \mathbb{C}\) be a repsentation of \(G^{\mathbb{C}}\) and \(E = P \times_{\rho} \mathbb{C}^l\) the associated vector bundle of \(P\). We define a bilinear form \(\eta\), a Hermitian metric \(g\) on \(E\) and an \({\rm End}(E)\)-valued 1-form \(\Phi\) by
\begin{align}
	&\rho(\Delta)(e_i) = \sum_{j=1}^{l}\eta(e_j,e_i)e_j,\ \ \ \rho(S^t \Delta)(e_i) = \sum_{j=1}^{l}g(e_j,e_i)e_j, \nonumber\\
	&\Phi = d\rho \circ \Phi^P, \nonumber
\end{align}
where \(\{e_j\}_{j=1}^{l}\) is the standard frame of \(E^{\pi} \simeq \mathbb{C}^* \times \mathbb{C}^l\). We can choose \(\rho\) so that \((E.\eta,g,\Phi)\) is isomorphic to the tt*-structure on \(E^{\mathbb{C}}_{k,N}\).
\begin{prop}\mbox{}\\
	\begin{itemize}
		\item [(1)] if \(\rho\) is a trivial representation \(\rho_0(g) = g\), then \((E,\eta,g,\Phi) \simeq (E^{\mathbb{C}P}_n,\eta^{\mathbb{C}P},g^{\mathbb{C}P},\Phi^{\mathbb{C}P})\).
		
		\item [(2)] if \(\rho = \bigwedge^k \rho_0\), then \((E,\eta,g,\Phi) \simeq (E^{\rm Gr}_{k,N},\eta^{\rm Gr},g^{\rm Gr},\Phi^{\rm Gr})\).
	\end{itemize}
\end{prop}
\begin{proof}
	(1) Obviously.\vskip\baselineskip
	
	\noindent (2) It follows from
	\begin{align}
		&\rho(g)(e_{r_1}\wedge \cdots \wedge e_{r_k}) = (ge_{r_1}) \wedge \cdots \wedge (ge_{r_k}), \nonumber\\
		&d\rho(X)(e_{r_1}\wedge \cdots \wedge e_{r_k}) = \sum_{j=1}^{k}e_{r_1} \wedge \cdots \wedge d\rho(X)(e_{r_j}) \wedge \cdots \wedge e_{r_k}. \nonumber
	\end{align}
\end{proof}
Hence, \((E^{\mathbb{C}P}_n,\eta^{\mathbb{C}P},g^{\mathbb{C}P},\Phi^{\mathbb{C}P})\) and \((E^{\rm Gr}_{k,N},\eta^{\rm Gr},g^{\rm Gr},\Phi^{\rm Gr})\) have the same tt*-structure as a principal \(G^{\mathbb{C}}\)-bundle \((P,c,\sigma,\Phi^P,A^P)\). In \cite{GH2017}, Guest, Ho described the tt*-Toda equation Lie-theoretically and it can be interpreted as the Hitchin's equation.

	\section*{Acknowledgement}
	This paper is a part of the outcome of research performed under a Waseda University Grant for Special Research Projects (Project number: 2025C-104).
	
	\section*{Conflict of interests}
	The author has no conflicts to disclose.

\bibliographystyle{plain}
\bibliography{mybibfile}

	\em
	\noindent
	Department of Applied Mathematics\newline
	Faculty of Science and Engineering\newline
	Waseda University\newline
	3-4-1 Okubo, Shinjuku, Tokyo 169-8555\newline
	JAPAN

\end{document}